\documentclass[
 reprint,
 amsmath,amssymb,
 aps]{revtex4-1}
\usepackage{comment}%
\usepackage{graphicx}
\usepackage{graphics}
\usepackage{dcolumn}
\usepackage{bm}

\usepackage[english]{varioref}

\begin{document}

\preprint{APS/123-QED}

\title{Ballooning, bulging and necking: an exact solution for longitudinal phase separation in elastic systems near a critical point.}

\author{Andrea Giudici}


\author{John S. Biggins}
\affiliation{Department of Engineering, University of Cambridge, Trumpington St., Cambridge CB21PZ, U.K.}%


\date{\today}

\begin{abstract}
Prominent examples of longitudinal phase separation in elastic systems include elastic necking, the propagation of a bulge in a cylindrical party balloon and the beading of a gel fiber subject to surface tension. Here we demonstrate that, if the parameters of such a system are tuned near a critical point (where the difference between the two phases vanishes) then the behaviour of all systems is given by the minimization of a simple and universal elastic energy familiar from Ginzburg-Landau theory in an external field. We  minimize this energy analytically, which yields not only the well known interfacial tanh solution, but also the complete set of stable and unstable solutions in both finite and infinite length systems, unveiling the elastic system's full shape evolution and hysteresis. Correspondingly, we also find analytic results for the the delay of onset, changes in criticality and ultimate suppression of instability with diminishing system length, demonstrating that our simple near-critical theory captures much of the complexity and choreography of far-from-critical systems. Finally, we find critical points for the three prominent examples of phase separation given above, and demonstrate how each system then follows the universal set of solutions.

\end{abstract}

\maketitle

\section{Introduction}
\begin{figure}[h]
\begin{center}
\includegraphics[width=0.9\columnwidth]{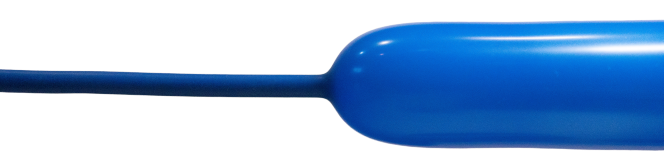}
\end{center}
\caption{\small Two phases of an inflated party balloon.}
\label{balloon}
\end{figure}
Imagine slowly inflating a cylindrical party balloon. At first, the balloon inflates homogeneously, but then, at a critical inflation, it suddenly separates along its length into a fat (more-inflated) bulged region and a thin less-inflated region \cite{mallock1891ii}. As inflation proceeds further (Fig. \ref{balloon})  the additional volume is not accommodated by radial expansion, but  by the fat region elongating (``propogating'') and consuming the thin region  \cite{ChaterHutchinson}. Once the thin region is totally consumed, the balloon is again homogeneous, and remains homogeneous during further inflation.

Ballooning is an example of an elastic instability, a highly topical category of phenomena where an elastic solid under load --- here inflation --- first deforms simply and then, past a  threshold loading, spontaneously adopts a more complicated shape. Classical instabilities, such as buckling \cite{euler1744additamentum, koiter1976current} and wrinkling \cite{allen1969, cerda2003geometry}, were studied as failure modes in stiff low-strain materials. However soft solids, such as elastomers, gels and many biological tissues, can sustain large strains, which generate a variety of new and dramatic geometrically motivated elastic instabilities, including ballooning \cite{Mallock1891,Hutchinson1985}, but also cavitation \cite{gent1959internal, ball1982discontinuous, cavitation}, sulcification \cite{biotbook,trujillo2008creasing, cai2012creasing, hohlfeld2011unfolding, dervaux2012mechanical, tallinen2013surface}   fingering  \cite{shull2000fingering, saintyves2013bulk, biggins2013digital, mora2014gravity, biggins2015fluid}, fringing  \cite{lin2016fringe, lin2017instabilities}, beading \cite{matsuo1992patterns,barriere1996peristaltic, mora2010capillarity, ciarletta2012peristaltic,ciarlettanonlinearjmps, ciarlettanonlinearpre, xuan2016finite,Xuan2017} and peristalsis \cite{xuan2016finite, cheewaruangroj2019peristaltic}. Moreover, soft high strain materials often survive instabilities,
allowing them to be harnessed by evolution to sculpt developing organs \cite{dervaux2011shape, ciarletta2012growth, kucken2004model, diab2013ruga, razavi2016surface, brainpnas, brain,loop, villi} and by engineers to underpin shape-shifting devices \cite{shim2012buckling, yang2015buckling, marthelot2017reversible, wang2018spatially}.

Ballooning may be familiar from daily life, but the same choreography, with one region consuming another, is also seen in several other 1D elastic instabilities including elastic necking \cite{antman1973nonuniqueness, ericksen1975equilibrium,Coleman1988}, strain-induced transformations in shape memory alloys \cite{bhattacharya2003microstructure}, liquid crystal elastomers \cite{Finkelmann1997}, the bend/snap of a builder's tape-measure \cite{Seffen1999}, and beading/bulging in gel fibers under surface tension \cite{ciarlettanonlinearpre, Xuan2017}. This choreography is characteristic of phase separation/coexistence, analogous to a  gas-liquid system in a piston which can accommodate large volume changes by switching material from liquid to gas \cite{Maxwell1875}. 

The simplest analysis of elastic phase separation considers the (zero-dimensional) energy for a homogeneous system \cite{knowles1978failure, Hutchinson1985}. Such treatment can identify the point of instability, characterise the two phases and find their length-fraction. However, this approach fails to resolve boundaries between the phases, including their shape, characteristic length, and energy-cost. In an infinite system split into two extensive phases, the boundary energy cost is  negligible, so the zero dimensional model captures the full hysteresis and choreography of the system. However, in finite systems boundaries fundamentally change the form of the instability \cite{Kyriakides1991}, altering its threshold, amplitude, and whether it is super- or sub-critical. 

At the opposite end of the spectrum, one can analyze a 1D elastic phase separation by minimizing the fully non-linear 3D elastic energy (e.g.\ \cite{Shi1996, Xuan2017}). This approach is guaranteed to capture the full form of the instability, but is typically only possible numerically. Between these two extremes, one can seek an effective 1D model, which allows a phase variable to vary continuously along the system, and augments the homogeneous energy by terms which penalize gradients to resolve the boundaries. This approach was pioneered in the van der Waals theory of fluid-vapour interfaces \cite{VanderWaals1979}. Within elasticity \cite{Coleman1988,triantafyllidis1986gradient,Triantafyllidis1993,Bardenhagen1994,Triantafyllidis1996} most such models are introduced heuristically, but regularising terms can also be obtained rigorously  from the 3D energy by dimensional reduction, as recently done for elastic rods \cite{Audoly2016,Xuan2017, Lestringant2020} and balloons \cite{Lestringant2018,Lestringant2020} under the assumption of small gradients and hence ``diffuse interfaces''. These 1D models are much more intelligible than their 3D counterparts, but they typically still only admit numerical solutions \cite{triantafyllidis1986gradient,Triantafyllidis1993,Lestringant2018}. Furthermore, they are derived by supposing small gradients, while still allowing the homogeneous energy to be minimized by radically different phases: the amplitude of the instability remains big while the gradients are small. As seen in Fig. \ref{balloon}, these assumptions are typically inconsistent since high amplitude implies a sharp interface and vice versa. 

Here, our approach is more limited: we focus on systems tuned near a ``critical-point'' in the energy landscape, where the difference between the two co-existing phases, and hence the amplitude of the instability, is truly small. In such a region, the interfaces are self-consistently diffuse, and expansion in the vicinity of the critical point leads to a simple and universal energy for near-critical 1D phase separation. The resultant energy is similar to near-critical van der Waals theory \cite{VanderWaals1979,Langer} as studied in the Cahn--Hillard equation \cite{Cahn1958,Novick-Cohen1984,Villain-Guillot2004}, but with the addition of a linear term from the elastic tension; equivalently it is similar to the Ginzburg-Landau magnetic energy \cite{vinokur2011ginzburg}  but in the presence of an external field. 

Minimizing this universal energy yields an Euler-Lagrange equation for the elastic body's shape. In elasticity, unlike thermodynamics, we are interested in both finite and infinite length systems, and in stable and unstable solution branches. Here, we obtain a complete analytic solution set and thus describe the system's full shape evolution and hysteresis. Although our theory is only valid near a critical point, it also delivers a rather complete portrait of high-amplitude systems, including resolved boundaries, and the changes of threshold and criticality with  diminishing length. In this paper, we first derive and solve our model, and then apply it to three examples: elastic necking, bulging under surface tension, and ballooning in cylindrical membranes.

\subsection{1D Model and Phase Separation}
\begin{figure}[h]
\begin{center}
\includegraphics[width=\columnwidth]{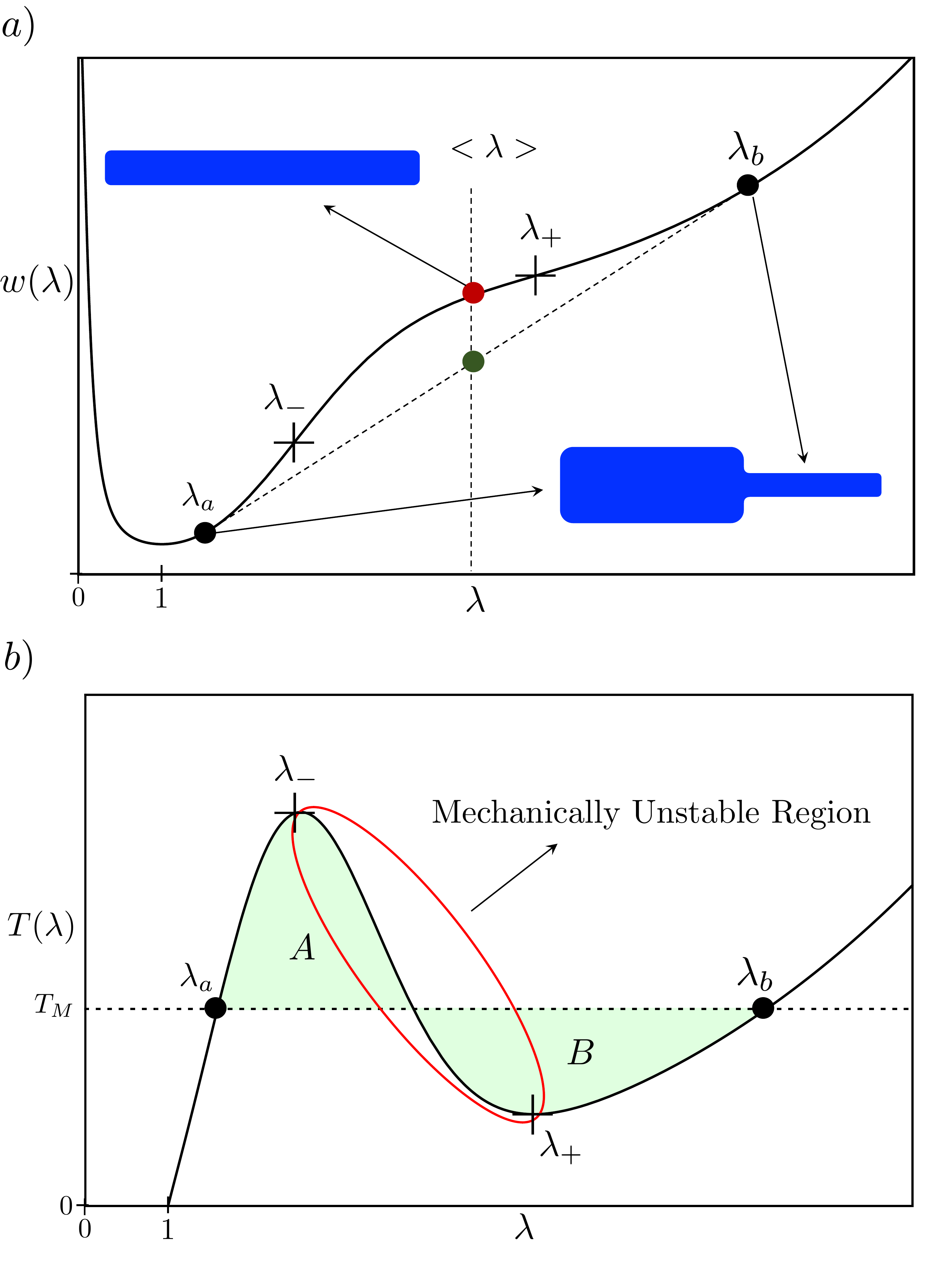}
\end{center}
\caption{\small (a) Example of locally concave (homogeneous) energy inducing phase separation and (b)  the corresponding tension curve, $T=\frac{\partial w}{\partial \lambda}$.}
\label{ps}
\end{figure}

We first recall the essentials of 1D elastic phase-separation in a simple context: a long elastic rod, which, when stretched by a factor of $\lambda$,  stores an elastic energy per unit length $w(\lambda)$ and hence bears a tension $T=\frac{\partial w}{\partial \lambda}$. If we impose an average stretch $<\lambda>$ the rod will adopt the configuration that minimizes the total elastic energy. Typically, $w(\lambda)$ is convex, but if $w(\lambda)$ has a concave region, and the average stretch of the system is brought into the concave region, the system can save energy by separating into two phases, $\lambda_a$ and $\lambda_b$, achieving an average intermediate stretch given by $<\lambda>=\nu \lambda_a+(1-\nu)\lambda_b$, where $\nu$ is the length fraction of the two phases, as shown in Figure \ref{ps} (a). This configuration is energetically favourable for, in the concave region, the chord connecting the phases lies below the original energy. As understood by Maxwell \cite{Maxwell1875}, optimal phase separation $(\lambda_a, \lambda_b)$ requires that the chord be tangent to the original energy at  $\lambda_a$ and $\lambda_b$, yielding the conditions,
\begin{align}
\frac{\partial w}{\partial \lambda}\bigg|_{\lambda_b}&=\frac{\partial w}{\partial \lambda}\bigg|_{\lambda_a}=T_M.
\label{maxwell2}\\
w(\lambda_b)&=w(\lambda_a)+\frac{\partial w}{\partial \lambda}\bigg|_{\lambda_a}(\lambda_b-\lambda_a)
\label{maxwell1}
\end{align}
where $T_M$ is known as the Maxwell tension. 

Figure \ref{ps} (b), which is a plot of the rod's tension $T(\lambda)=\frac{\partial w}{\partial \lambda}$, provides an alternative perspective. Concavity in $w(\lambda)$ produces an unstable region where tension decreases with length. Phase separation is triggered upon entry into this region, at $\lambda_{\pm}$, which are the limits of concavity ($w''(\lambda{_\pm})=0$) and known as the instability (or Consid\'ere) points. During phase separation, the system settles down to $T_M$ and associated phases $\lambda_a$ and $\lambda_b$. These can be found graphically from Figure \ref{ps} b) by applying the celebrated equal-area rule for $A$ and $B$; equivalent to the above common-tangent condition.

\begin{figure}[h]
\begin{center}
\includegraphics[width=\columnwidth]{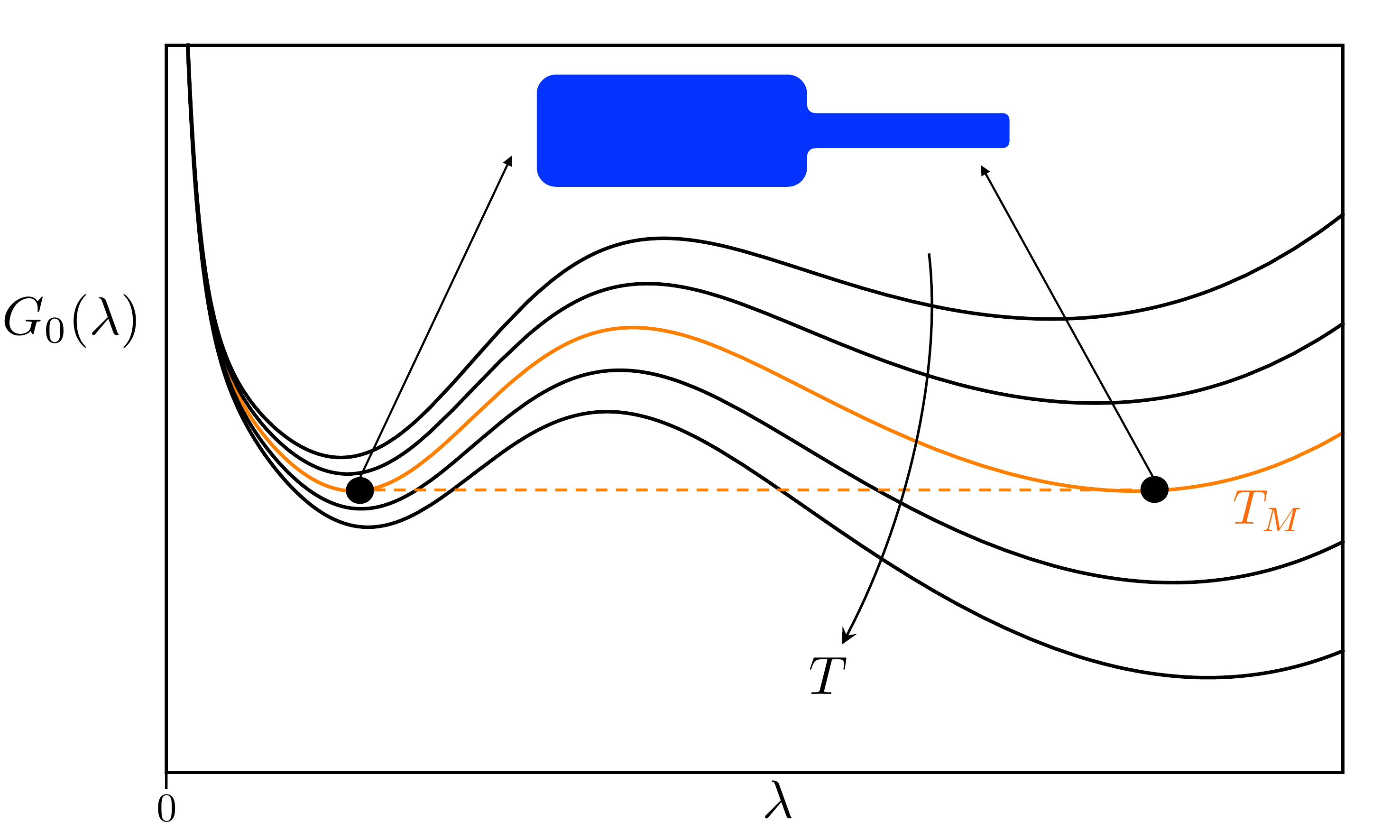}
\end{center}
\caption{\small Plot of the (homogeneous) Gibbs energy $G_0(\lambda,T)=w_0(\lambda)-T\lambda$ at different values of $T$. The curve at $T=T_M$, shown in orange, has the the two minima being energetically equivalent.}
\label{gibbs}
\end{figure}

Finally, phase separation can also be understood by writing the total homogeneous energy (Gibbs energy) as
\begin{align}
G(\lambda,T)=w(\lambda) - T \lambda
\label{potenh}
\end{align}
where the second term arises as a Lagrange multiplier constraining the total stretch. At equilibrium, $\frac{\partial G}{\partial \lambda}=0$ which implies $T=\frac{\partial w}{\partial \lambda}$ and hence the Lagrange multiplier is simply the tension. As shown in Fig.\ \ref{gibbs}, as $T$ traverses $T_M$, the global minimum of $G$ jumps from $\lambda_a$ to $\lambda_b$.

Regrettably, the homogeneous analysis cannot comment on the boundaries between phases. These boundaries can be resolved by considering the full 3D displacement $\mathbf{u}$ and corresponding elastic energy density $W(\nabla \mathbf{u})$. The full form of the system is given by the 3D displacement field that minimizes $\mathcal{E}=\int W \mathrm{d}V$ subject to either fixed displacement boundary conditions at the ends or fixed load \cite{Pamplona2006}. However, although such energies can be minimised in finite elements, we do not know of any analytic minimizations in the context of 1D phase separation. 

An intermediate approach is to minimise the 3D energy subject to the constraint that the centre-line stretch follows the function $\lambda(Z)$ \cite{Lestringant2020}. This produces a 1D energy density $w(\lambda,\lambda', \lambda''...)$ which depends only on $\lambda$ and its $Z$ derivatives. If one also assumes that $\lambda$ is slowly varying, one can Taylor expand in small derivatives to obtain a regularized 1D energy
\begin{equation}
\mathcal{E}=\int_L\left(w_0(\lambda)+\frac{1}{2}B(\lambda) \lambda'^2\right) dZ
\end{equation}
where $L$ is the total length in the reference configuration. The first term, $w_0(\lambda)$ has gained a $0$ subscript to identify it as the energy of the homogeneous system, although its argument is now a spatially varying function $\lambda(Z)$. We shall repeatedly use this subscript to indicate functions that naturally arise in the discussion of a homogeneous system. The second term in the integral, often referred to as the \textit{regularising term}, penalises gradients in $\lambda$ and resolves the phase boundary \cite{Audoly2016, Xuan2017,Lestringant2018,triantafyllidis1986gradient,Triantafyllidis1993}.  In principle, one now minimises variationally over $\lambda(Z)$ to find the form of the rod, although it is still typically only possible to conduct this final minimization numerically \cite{Lestringant2018,Coleman1988,triantafyllidis1986gradient,Triantafyllidis1993}.

\section{Near critical phase separation}
\begin{figure}[h]
\begin{center}
\includegraphics[width=\columnwidth]{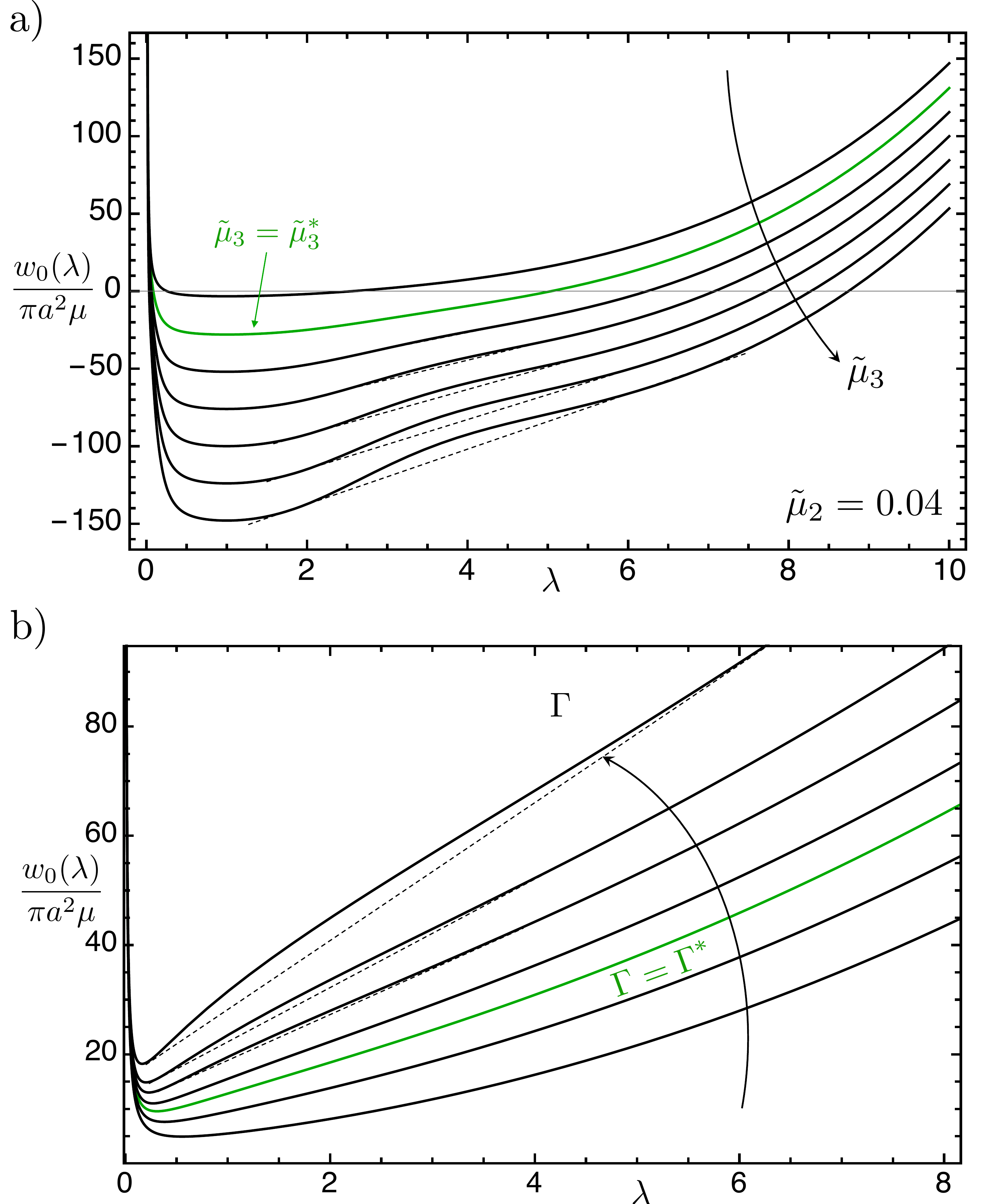}
\caption{\small  (a) Coleman-Newman energy of a stretched polymeric fiber (eqn.\ \eqref{eqn:coleman}) tuned by a constitutive parameter $\tilde{\mu}_3$.  (b) Stretch energy of a neo-Hookean fiber subject to surface tension tuned via the elastocapillary number $\Gamma$ (eqn. \eqref{rodsurfaceen}). Both become convex beyond a tuning parameter threshold.}\label{figurehm}
\end{center}
\end{figure}

In this paper we give a complete analytic solution to the problem by further assuming that the amplitude difference between the  phases is small. This assumption is more limited than small gradients, but has the advantage of being self consistent, since small amplitude generates small gradients and vice versa.  Such a region arises naturally if the system has an additional parameter $\xi$, which is fixed during a given phase-separation experiment, but controls the width of the concavity in $w_0(\lambda; \xi)$, with concavity completely disappearing below the critical value $\xi^*$. As shown for two examples in Figure \ref{figurehm}, $\xi$ is typically a parameter that specifies the magnitude of some concavity-inducing physical effect (e.g.\ surface tension), and $\xi^*$ is the threshold value at which the new concave physics overwhelms the naive elastic convexity. 

By tuning $\xi$ a small distance above the critical point, $\xi=\xi^*+\varepsilon$, we re-introduce a slight concavity into the energy, which leads to a slowly-varying ($\lambda'$ small),  low amplitude ($\lambda_b-\lambda_a$ small) phase separation. Expanding $G$ in these small quantities leads to a universal form, which we minimise analytically to find a universal set of inhomogeneous phase-separation solutions.

\subsection{Critical Point Expansion}
Given a regularized 1D elastic energy of the form appropriate to describe necking in an elastic rod,
\begin{equation}
w(\lambda, \lambda'; \xi)=w_0(\lambda;\xi)+\frac{1}{2} B(\lambda;\xi) \lambda'^2 \label{eqn:wdef}
\end{equation}
and an associated regularized Gibbs energy
\begin{align}
G(\lambda, \lambda',T; \xi)&=w(\lambda, \lambda';\xi)-T\lambda\label{eqg:Gdeff}\\
&\equiv G_0(\lambda, T;\xi)+\frac{1}{2} B(\lambda;\xi) \lambda'^2 \notag,
\end{align}
 our first task is to find the critical point, $(\lambda^*,T^*;\xi^*)$, for the onset of concavity in the homogeneous problem. The critical parameter $\xi^*$ corresponds to the merger of the Considére points ($\lambda{_\pm}$), giving an inflection point of the tension and the disappearance of the unstable region. Thus, to find  $(\lambda^*,T^*;\xi^*)$ we solve
 \begin{equation}
     w_{\lambda\lambda}^*=w_{\lambda\lambda\lambda}^*=0 ; \hspace{3pt} w_\lambda^*=T^*
\label{criticalq}
 \end{equation}
where $*$ denotes evaluation at the critical point, and the subscripts are derivatives. Note that these conditions are really applied to $w_0$, but since the system remains homogeneous at the critical point, we have $w_0^*=w^*$ and the same holds for all derivatives other than $\lambda'$. For simplicity, we will only write the subscript $0$ when necessary. 
 
We now set $\xi=\xi^*+\varepsilon$ and expand around the critical point. We first focus on $w_0$ and find the near critical Consid\'ere stretches ($w_{0,\lambda\lambda}=0$) to characterize the width of the concavity. Expanding this equation and  keeping the leading term in $\lambda-\lambda^*$ and $\varepsilon$ gives
\begin{equation}
w_{0\lambda \lambda}(\lambda;\xi)=w_{\lambda \lambda \xi}^* \varepsilon+\frac{1}{2}w_{\lambda\lambda\lambda\lambda}^* (\lambda-\lambda^*)^2+...=0.
\end{equation}
Therefore, the near critical Considére stretches are at $
\lambda_{\pm} =\lambda^* \pm\sqrt{2w_{\lambda \lambda \xi}^* \varepsilon/w_{\lambda\lambda\lambda\lambda}^*}+...\,$ and the width of the concavity (amplitude) scales as $\delta\lambda=\lambda-\lambda^* \sim \sqrt{\varepsilon}$.

We now expand $w_0$ to order $\varepsilon^2$ about the critical point, 
\begin{align}
w_0(&\lambda;\xi)=w_0(\lambda^*;\xi)\\&
+(w_{\lambda \xi}^*  \varepsilon+T^*)\delta \lambda +\frac{1}{2}w_{\lambda \lambda \xi}^*  \varepsilon\delta \lambda ^2+\frac{1}{24}w_{\lambda\lambda\lambda\lambda}^* \delta\lambda^4+...\notag
\end{align}
where we have again simplified using eqn. \eqref{criticalq}. Next, expanding $G_0$ requires us to expand $\lambda T=T\lambda^*+T\delta \lambda$. By definition, the minima of $G_0$ are equally deep at the Maxwell tension, which means that the linear $\delta \lambda$ term in $G_0$ must vanish, requiring  $T_M=T^*+w_{\lambda \xi}^*\varepsilon$. Finally, writing $T=T_M(\xi)+\delta T$, we get the full expansion of $G$,
\begin{align}
G(\lambda,&\lambda',T;\xi)=G(\lambda^*,0,T_M;\xi^*)+ \label{GF} \\ & +\frac{1}{2}w_{\lambda \lambda \xi}^*  \varepsilon\delta \lambda ^2+
\frac{1}{24}w_{\lambda\lambda\lambda\lambda}^* \delta\lambda^4 
+  
\frac{1}{2} B^*\delta\lambda'^2-\delta T \delta\lambda, \notag
\end{align}
which corresponds to Ginzburg-Landau magnetism in an external field. At the Maxwell tension ($\delta T=0$) the homogeneous energy is a symmetric quartic with minima at $\delta \lambda_{ab}=\pm\sqrt{-\frac{6\,w_{\lambda \lambda \xi}^*\varepsilon}{w_{\lambda\lambda\lambda\lambda}^*}}$, confirming that amplitude scales as $\delta \lambda \sim \sqrt{\varepsilon}$ and vanishes at the critical point.

\subsection{Non Homogeneous Solutions}
 \begin{figure}[b]
\begin{center}
\includegraphics[width=\columnwidth]{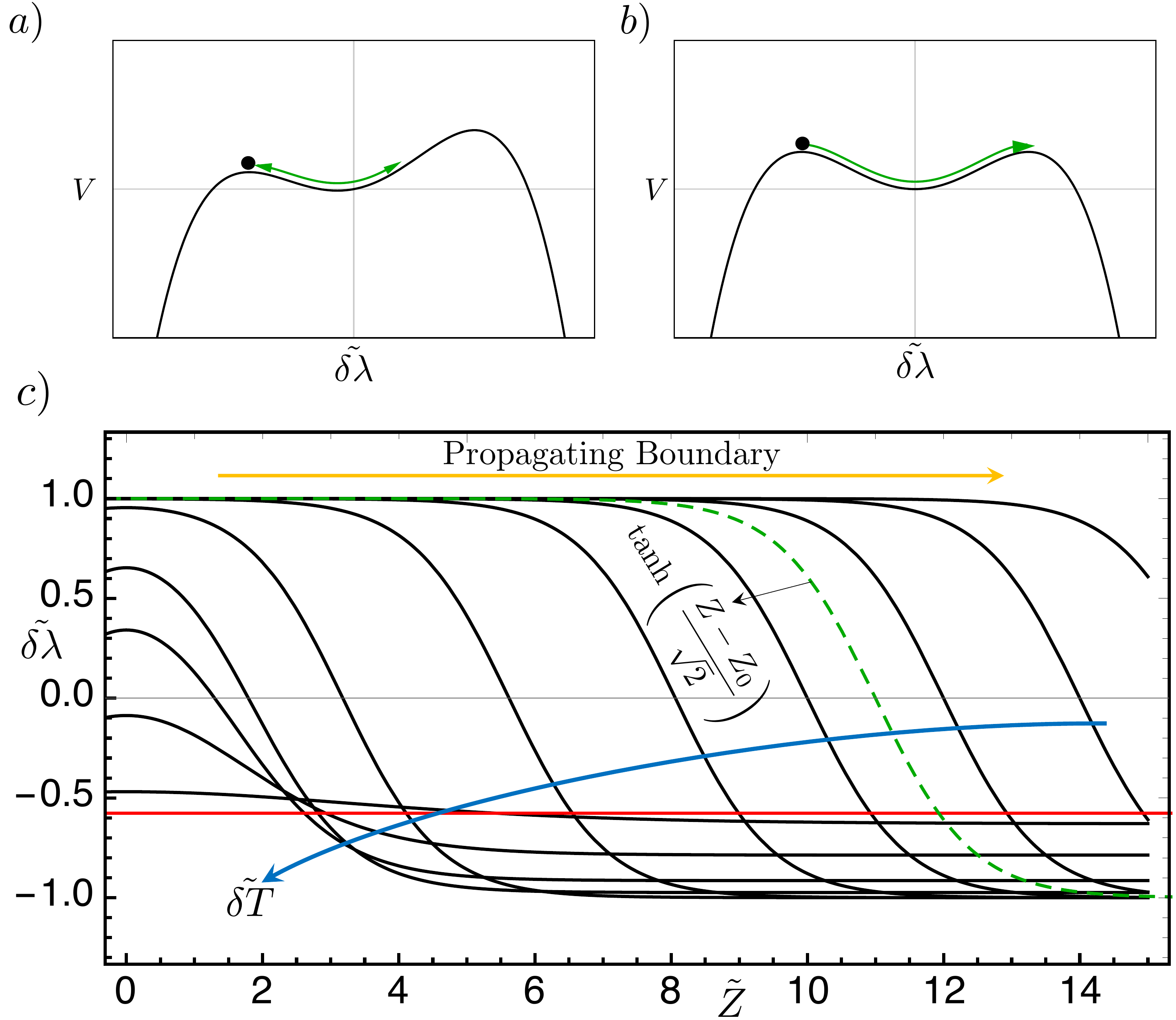}
\end{center}
\caption{ \small(a-b) Examples of motion when $\tilde{\delta T}>0$ (a) and $\tilde{\delta T}=0$ (b).
(c) Plot of (half) solutions for long systems at different $\tilde{\delta T}$. In dashed green, we compare the shape for $\tilde{\delta T}$ approaching $0$ to the propagating hyper-tangent solution.}
\label{figsolitonsol}
\end{figure}
The expansion of $G$ reveals a universal functional dependence on $\delta \lambda$, and its minimization will yield a universal $\delta \lambda(Z)$. First, we nondimensionalize by writing:
\begin{align}
&\delta \lambda= \sqrt{\frac{-6\,w_{\lambda \lambda \xi}^*\varepsilon}{ w_{\lambda\lambda\lambda\lambda}^*}} \tilde{\delta \lambda}, \hspace{6pt}
Z=\sqrt{\frac{ -B^*}{w_{\lambda \lambda \xi}^*\varepsilon}}\tilde{Z} \notag \\ &\delta T=\sqrt{\frac{-6 (w_{\lambda \lambda \xi}\varepsilon)^3}{w_{\lambda\lambda\lambda\lambda}^*}}\tilde{\delta{T}}.
\label{rescale}
\end{align}
These rescalings also absorb the near-critical  $\sqrt{\varepsilon}$ scaling of amplitude ($\delta \lambda$), the $1/\sqrt{\varepsilon}$ scaling of the characteristic lengthscale ($\delta \lambda/\delta \lambda'$), and the $\varepsilon^{3/2}$ scaling of $\delta T$ required to span the instability. Having accounted for all the $\varepsilon$ dependence, the re-scaled Gibbs energy is $\varepsilon$ independent:
\begin{align}
\tilde{G}&=\frac{w_{\lambda \lambda \lambda \lambda}}{6 (\varepsilon w_{\lambda\lambda\xi}^*)^2}\left(G(\lambda,\lambda',T;\xi)-G(\lambda^*,0,T_M;\xi^*)\right)\notag\\
&=\frac{1}{2}\tilde{\delta \lambda}'(\tilde{Z})^2-\tilde{\delta T} \tilde{\delta \lambda}(\tilde{Z}) -\frac{1}{2}\tilde{\delta \lambda}(\tilde{Z})^2+\frac{1}{4}\tilde{\delta \lambda}(\tilde{Z})^4.
\label{Gb}
\end{align}
The homogeneous part, $\tilde{G_0}$, is a symmetric quartic at $\tilde{\delta T}=0$, with rescaled minima at $\tilde{\delta \lambda}_{ab}=\pm1$, and $\tilde{G_0}$ is single-welled if $|\tilde{\delta T}|> \frac{2}{3 \sqrt{3}}$. Minimising with respect to variations in $\tilde{\delta \lambda}$ gives a nonlinear differential equation:
\begin{equation}
\tilde{\delta \lambda}''(\tilde{Z})=-\delta T-\tilde{\delta \lambda}(\tilde{Z})+\tilde{\delta \lambda}(\tilde{Z})^3,
\label{diffeq}
\end{equation}
augmented by the natural boundary conditions
\begin{equation}
\tilde{\delta \lambda}'\big|_{\frac{\tilde{L}}{2}}=\tilde{\delta \lambda}'\big|_{-\frac{\tilde{L}}{2}}=0.
\label{b.c.}
\end{equation}
The solution to \eqref{diffeq} and \eqref{b.c.} completely defines the deformation of the elastic body. 
To build intuition, we note that if we interpret $\tilde{Z}$ as time and $\tilde{\delta \lambda}$ as position,  equation \eqref{diffeq} describes the motion of a unit mass particle in an upside down quartic potential $V(\tilde{\delta \lambda})=\delta T \tilde{\delta \lambda}+\frac{1}{2}\tilde{\delta \lambda}^2-\frac{1}{4}\tilde{\delta \lambda}^4=-\tilde{G_0}(\tilde{\delta \lambda})$, and the boundary conditions require the particle to be stationary at the start and end of the motion. One can thus directly find a first integral equivalent to conservation of energy:
\begin{equation}
\frac{1}{2}\tilde{\delta \lambda}'(\tilde{Z})^2+\delta T \tilde{\delta \lambda}(\tilde{Z}) +\frac{1}{2}\tilde{\delta \lambda}(\tilde{Z})^2-\frac{1}{4}\tilde{\delta \lambda}(\tilde{Z})^4=E.
\label{firstintegral}
\end{equation}
We first focus on infinite length (time) solutions. Eqn.\ \eqref{firstintegral} clearly admits homogeneous solutions if the system sits at a minimum of $\tilde{G_0}$. At the Maxwell tension ($\tilde{\delta T} = 0$),
there is also a solution in which the particle slides from one maxima of $V$ to the other (Fig.\ \ref{figsolitonsol}b):
\begin{equation}
\tilde{\delta \lambda}(\tilde{Z})=\tanh \left(\frac{\tilde{Z}-Z_0}{\sqrt{2}} \right).
\label{tanh}
\end{equation}
 This solution describes the phase boundary, and is a classic result familiar from near-critical van der Waals theory \cite{VanderWaals1979,Langer,Cahn1958,Novick-Cohen1984,Villain-Guillot2004} and Ginzburg-Landau magnetic domain walls. 
 
\begin{figure}[b]
\begin{center}
\includegraphics[width=\columnwidth]{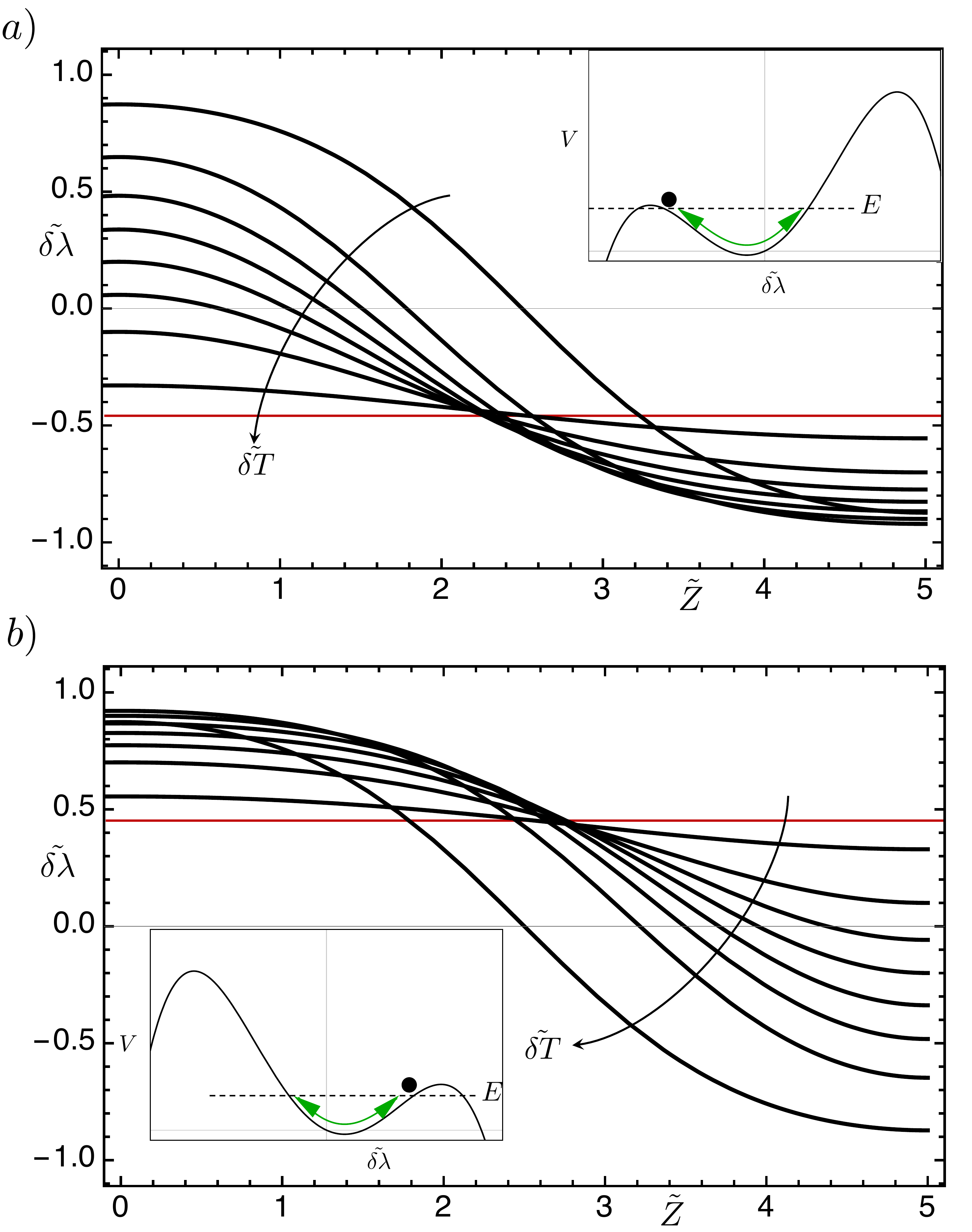}
\end{center}
\caption{\small Plots of $\tilde{\delta \lambda}(\tilde{Z})$ for a system of length $\tilde{L}=10$ at different values of $\tilde{\delta T}$, a) for $\tilde{\delta T}\geq0$ and b) for $\tilde{\delta T}\leq 0$. The two have been separated to avoid overlapping of solutions as well as to show (in the insets) the type of motion described. Note, this is half of the bulge/neck solution, the other half is, by symmetry, just a reflection about the vertical axis.}
\label{fl}
\end{figure}

Away from the Maxwell tension the quartic is asymmetric, but one still has infinite length solutions in which the particle starts and returns to the lowest maxima of $V$ (Fig.\ \ref{figsolitonsol}a), which is at $\tilde{\delta \lambda}_{max}$.  These solutions are:
\begin{align}
&\tilde{\delta \lambda}(\tilde{Z})=\tilde{\delta \lambda}_{max}+\frac{6 A}{2 B+\sqrt{4 B^2-18 A } \cosh \left(\sqrt{A} (\tilde{Z}-\tilde{Z}_0)\right)}.\notag \\
&\mathrm{with\ \ \ \ }A=-1+3\tilde{\delta \lambda}_{max}^2\,,\,\,\,\,\,\,\,B=-3 \tilde{\delta \lambda}_{max},\label{AB}
\end{align}
as is familiar from the motion of an electron confined in a asymmetric quartic potential \cite{Gordon1988}. The resultant family of solutions for increasing $\delta T>0$ above the Maxwell tension are shown in Figure \ref{figsolitonsol} (c). As expected, when $\tilde{\delta T}\gtrsim 0$, the solution very closely resembles a propagating phase separated $\tanh$ solution, but as $\tilde{\delta T} \to \frac{2}{3 \sqrt{3}}$ (i.e. as $T$ approaches its Consid\'ere value) the central phase vanishes and the system becomes homogeneous. 

\begin{figure*}[t]
\begin{center}
\includegraphics[width=\textwidth]{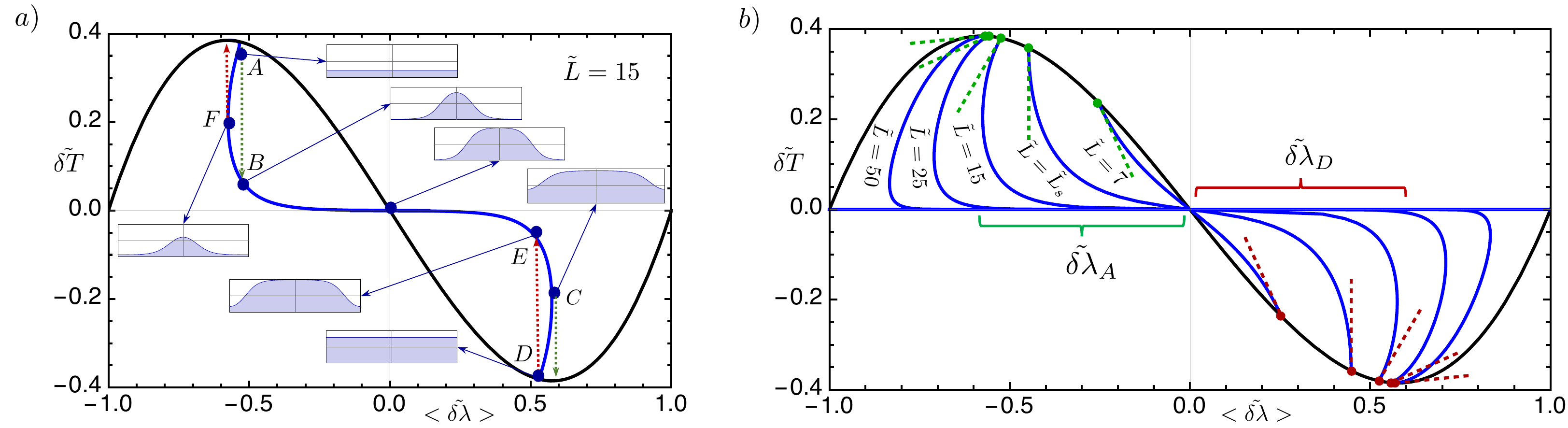}
\end{center}
\caption{\small a) Hysteresis loop for a system of length $\tilde{L}=15>\tilde{L}_s$ with the forward path ($A$ - $B$ - $C$) being different from the backwards path ($D$ - $E$ - $F$). b) Plot of bifurcated branches, instability points and initial gradients for systems of different length. The bifurcation points $\tilde{\delta \lambda}_A$ and $\tilde{\delta \lambda}_D$ correspond to the instability points in $A$ and $D$ in a).}
\label{f7} 
\end{figure*}

In finite length systems, one needs solutions in which the particle conducts finite period oscillations in the central well of $V$ (Fig. \ref{fl}). In this situation, we may write $E-V(\tilde{\delta \lambda})=\frac{1}{4}(\tilde{\delta \lambda}-\tilde{\delta \lambda}_1)(\tilde{\delta \lambda}-\tilde{\delta \lambda}_2)(\tilde{\delta \lambda}-\tilde{\delta \lambda}_3)(\tilde{\delta \lambda}-\tilde{\delta \lambda}_4)$ where the $\tilde{\delta \lambda}_is$ are the four ordered roots of the polynomial $V-E$, with $\tilde{\delta\lambda}_2$ and $\tilde{\delta \lambda}_3$ bounding the central well. The quartic polynomial $V-E$ is \emph{depressed} (no cubic term) so exact solutions for the $\delta \lambda_i$ in terms of $\tilde{\delta T}$ and $E$ are easily obtained (e.g.\ in \textit{Mathematica}).  Letting $\Delta_{ij}=\tilde{\delta \lambda_i}-\tilde{\delta \lambda_j}$, and integrating \eqref{firstintegral} a second time, we can write the solution as:
\begin{equation}
\tilde{\delta \lambda}(\tilde{Z})=\frac{\tilde{\delta \lambda}_2\Delta_{41}-\tilde{\delta \lambda}_1\Delta_{42}\phi^2(\tilde{Z}-\tilde{Z}_0)}{\Delta_{41}-\Delta_{42}\phi^2(\tilde{Z}-\tilde{Z}_0)}\label{eqn:finitelengthsolution}
\end{equation}
where
\begin{equation}
\small \phi(\tilde{Z})=\text{sn}\left(\frac{\tilde{Z}\sqrt{\Delta_{32}\cdot \Delta_{41}}}{2 \sqrt{2}}\bigg|\frac{\Delta_{31}\cdot \Delta_{42}}{\Delta_{32}\cdot \Delta_{41}}\right) \label{eqn:eliptic}
\end{equation}
and $\text{sn}(u|m)$ is the Jacobi elliptic function. 

The half length of the system is then given by the inverse of \eqref{eqn:finitelengthsolution} evaluated between $\tilde{\delta \lambda}_3$ and $\tilde{\delta \lambda}_2$. Thus, the total length is given by
\begin{equation}
\tilde{L}= \sqrt{\frac{32}{\Delta_{41}\cdot \Delta_{32}}} F\left(\sin
   ^{-1}\left(\sqrt{\frac{\Delta_{32} \cdot \Delta_{41}}{\Delta_{31}\cdot \Delta_{42}}}\right)|\frac{\Delta_{31}\cdot \Delta_{42}}{\Delta_{32} \cdot \Delta_{41}}\right)
   \label{eqn:length}
\end{equation}
where $F(u|v)$ is the elliptic integral of the first kind. Note that once $\tilde{\delta T}$ is fixed, the length $\tilde{L}$ only depends on $E$. Thus, at any tension, we can solve \eqref{eqn:length} to find the appropriate $E$ which yields solutions of length $\tilde{L}$. For every length $\tilde{L}$ we therefore have a 1D family of shapes, paramaterised by $\tilde{\delta T}$, that determines the whole evolution of a bulge/neck. An example of the shapes of a bulge/neck for a system with $\tilde{L}=10$ is shown in Fig.\ \ref{fl}.

\subsection{Length dependence of threshold and hysteresis}

In an experiment, one typically fixes $<\lambda>$ by clamping the ends, rather than directly imposing $T$; consequently many of these solutions may in fact be unstable and unobserved. Working backwards, for every choice of $\tilde{\delta T}$, there is a shape characterising the equilibrium state which has an associated average stretch given by
\begin{align}
<\lambda>&=\lambda^*+\frac{1}{\tilde{L}}\int_{-\tilde{L}/2}^{\tilde{L}/2}\tilde{\delta \lambda}(\tilde{Z})\,d\tilde{Z}.
\label{totstretch}
\end{align}
We can thus plot the family of inhomogeneous and homogeneous solutions as paths in $T$ - $\lambda$ to understand the choreography.  In long systems during loading, the inhomogenous branch emerges close to the Consid\'ere point, $(\tilde{\delta \lambda}_{-},\tilde{\delta T}_{-})=(-\frac{1}{\sqrt{3}},+\frac{2}{3 \sqrt{3}})$,  with a backward gradient, indicating a sub-critical instability with a hysteresis loop, as shown in Fig.\ \ref{f7}a. However, in shorter systems (Fig. \ref{f7}b)  the true instability, $(\tilde{\delta \lambda}_{A},\tilde{\delta T}_{A})$, moves away from the Consid\'ere point and towards the origin, and ultimately annihilates with the unloading instability point ($\tilde{\delta \lambda}_D=-\tilde{\delta \lambda}_A$) at a critical length, $\tilde{L}_h$, below which there are no inhomogeneous solutions. Expanding the solution to first order around $(\tilde{\delta \lambda}_{A},\tilde{\delta T}_{A})$ (appendix A: linear stability analysis) we can identify the true thresholds as:
\begin{equation}
\tilde{\delta \lambda}_{A}=-\frac{\sqrt{\tilde{L}^2-4 \pi ^2}}{\sqrt{3} \tilde{L}}\,\,\, \implies \,\,\,\,\, \tilde{L}_h=2 \pi.
\end{equation}
Furthermore, in  Fig.\ \ref{f7}b the initial gradient of the inhomogeneous branch, $m_A$, changes sign, indicating a change from sub- to super-critical behaviour at length $\tilde{L}_s>\tilde{L}_h$. Expanding to higher order around the instability point (appendix A, Koiter analysis \cite{koiter1970stability}) we find: 
\begin{equation}
m_{A}=\frac{20 \pi ^2 \tilde{L}^2-56 \pi ^4}{\tilde{L}^4-10 \pi ^2 \tilde{L}^2} \,\,\, \implies\,\,\, \tilde{L}_s=\sqrt{10} \pi.
\end{equation}

\begin{figure}[h]
\begin{center}
\includegraphics[width=\columnwidth]{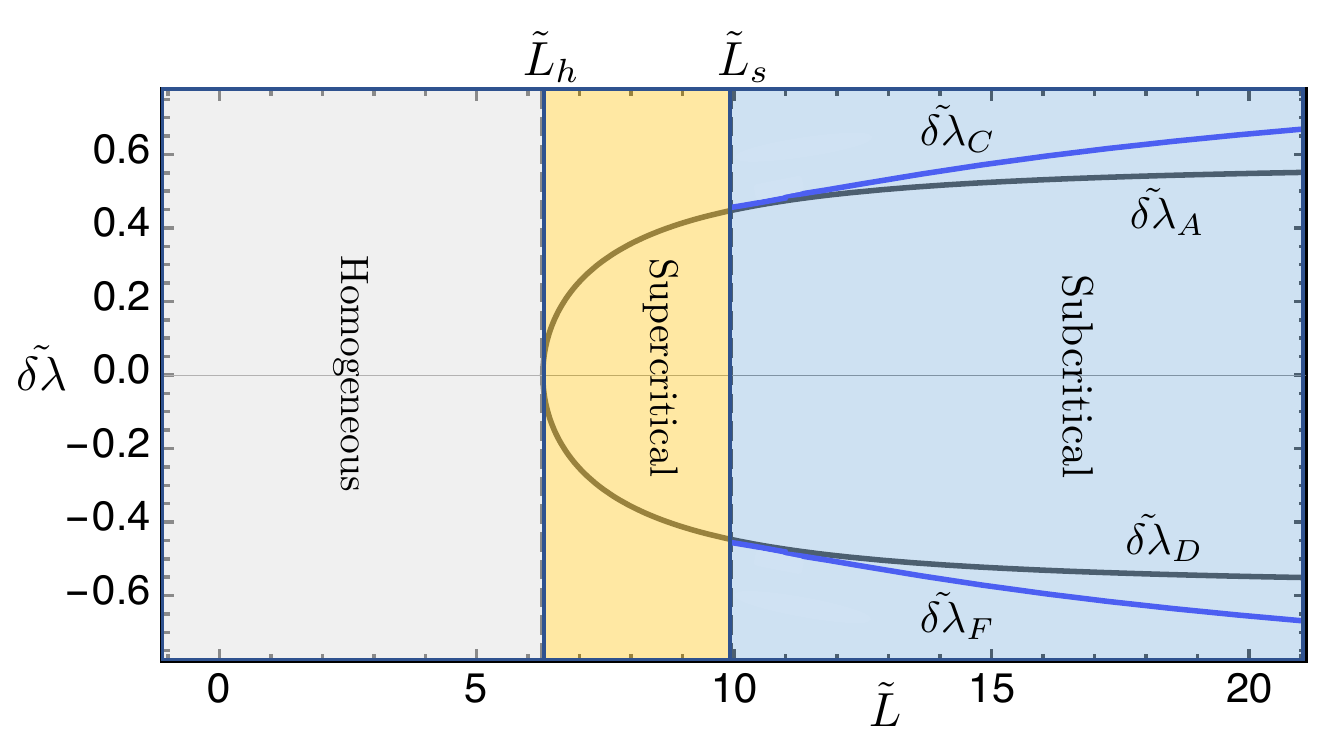}
\end{center}
\caption{\small Length dependent phase separation behaviour showing regions of homogeneous, super- and sub -critical behaviour. The lines map the hysteresis points $A$, $C$ , $D$ and $F$ in Fig. \ref{f7} b), which determine the different types of instability.}
\label{f8}
\end{figure}

This length dependence is summarised in Figure \ref{f8} showing the three regions of homogeneous, supercritical and subcritical behaviour. It is consistent with the behaviour of systems far from the critical point highlighted by previous authors 
\cite{Kyriakides1991,ciarlettanonlinearpre, Lestringant2018}, with the presences of subcritical instabilities and hysteresis in long systems being suppressed in shorter ones. Our analytic results for these threshold lengths, alongside our general analytic solutions, thus offer not only a complete description of any near-critical 1-D elastic phase separation, but also a satisfactory miniature portrait of high amplitude behaviour.

\section{Examples}
We present three different examples of near critical 1D phase separation in elastic systems. The first two examples are both elastic rods phase-separating under longitudinal stretch,  with phase separation being driven by intrinsic material concavity in the hyper-elastic energy in the first case, and by an additional surface tension in the second. The final example is phase separation in a cylindrical balloon. In each case, our example builds on recent work developing small gradient theories for elastic rods \cite{Audoly2016}, rods with surface tension \cite{Xuan2017} and for cylindrical balloons \cite{Lestringant2018}. 

\subsection{Necking in Polymeric Fibers Under Tension}
Solid bars are often observed to neck under extension. In most stiff materials, although the onset of necking is associated with a concavity (softening) in the elastic energy, the neck that forms is dominated by plasticity and lies outside the realm of this work \cite{chen1971necking,Audoly2019}. Polymeric systems form necks as well, as first described by Carothers and Hill in 1932 \cite{carothers1932`}, some of which are deemed to be purely elastic. For example, Coleman and Newman \cite{Coleman1988} studied elastic necking during the cold-drawing of nylon fibers, and proposed a suitable one dimensional homogeneous energy function,
\begin{align}
\notag \frac{w_0(\lambda;\tilde{\mu}_3)}{\mu \pi a^2 }=&\frac{1}{2}(\lambda^2+2 \lambda^{-1})+\frac{\tilde{\mu}_2}{4}(\lambda^4+2 \lambda^{-2})\\
\notag &-2 \tilde{\mu}_3(\lambda^2+3)e^{(-1/4(\lambda^2-1))}\\
&-4 \tilde{\mu}_3 (\lambda^{-1}+3)e^{(-1/4(\lambda^{-1}-1))},
\label{eqn:coleman}
\end{align}
where $\mu$ is the small strain shear modulus,  $\tilde{\mu}_2$ and $\tilde{\mu}_3$ are dimensionless material parameters, and $a$ is the rod radius. The first term in the above energy describes basic (neo-Hookean) elastic behaviour; the second term captures the elastomer stiffening at high strains, and the $\tilde{\mu}_3$ terms (which are responsible for introducing the concavity) represent a phenomenological attempt to capture ``the loss of structural and rotational stability'' when the material is forced into alignment.

\begin{figure}[b]
\begin{center}
\includegraphics[width=\columnwidth]{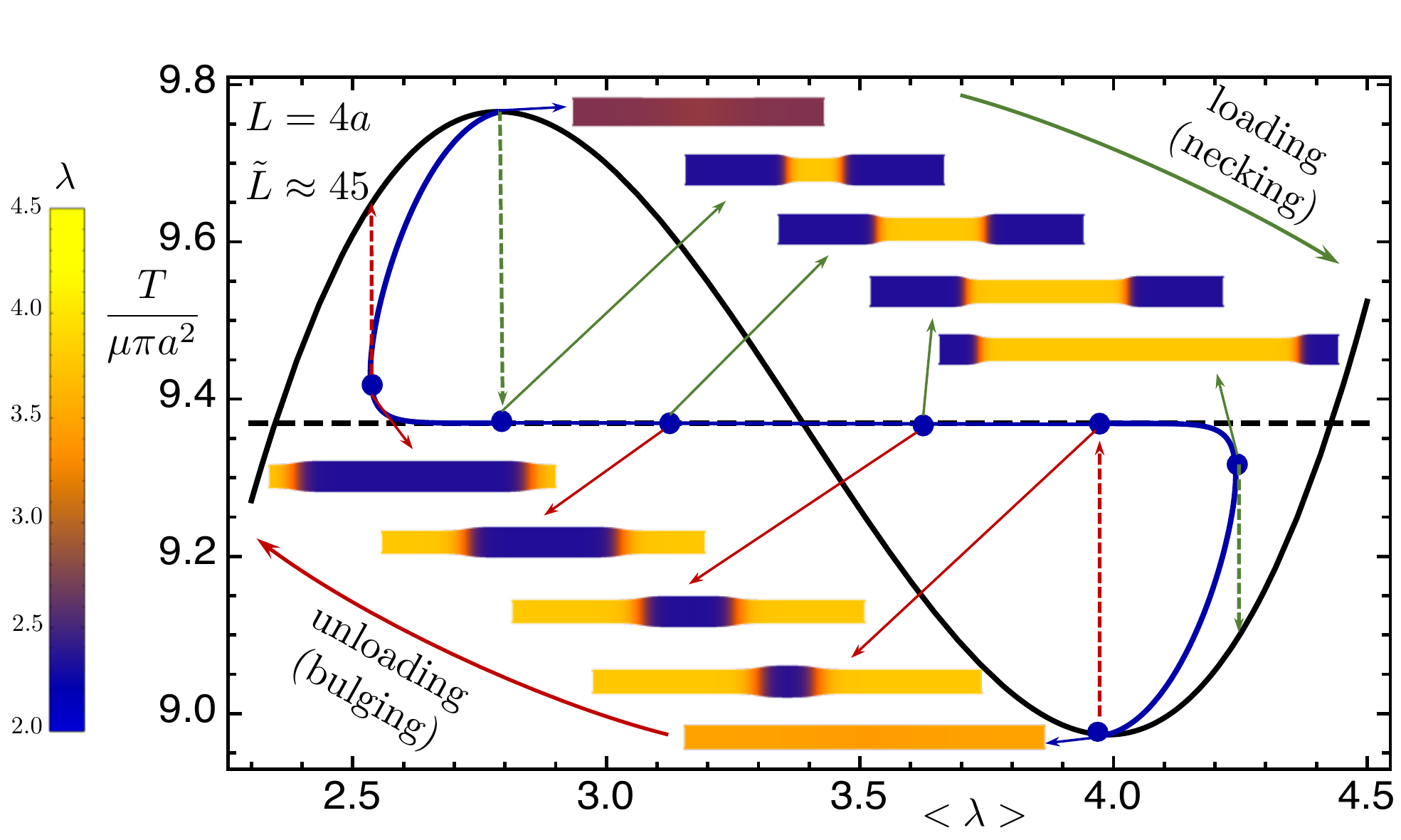}
\end{center}
\caption{ \small  Hysteresis and shape of a rod under tension, following the Coleman-Newman elastic energy with $\epsilon=\tilde{\mu}-\tilde{\mu_3}=0.5$.}
\label{neck}
\end{figure}

Given any $\tilde{\mu}_2$,  it is possible to find $\tilde{\mu}_3$ such that $w_0(\lambda)$ has a critical point satisfying eqn.\eqref{criticalq}; following the original paper, we fix $\tilde{\mu}_2=0.04$, and take $\xi=\tilde{\mu}_3$ as our dimensionless tuning parameter. The change of the concavity of the homogeneous energy as a function of $\tilde{\mu}_3$ is shown in Figure \ref{figurehm} (a) and corresponds to the critical point at $(\lambda^*,\tilde{\mu}_3^*)=(3.388...,1.228...)$.  
Audoly and Hutchinson \cite{Audoly2016} have showed that it is possible to find a regularized energy for prismatic orthotropic elastic systems through dimensional reduction. In the case of a cylindrical rod, one obtains a regularization term $\frac{1}{2}\frac{a^2 }{8 \lambda^4}\frac{d w_0}{d\lambda} \lambda'^2$ which leads to the regularized energy:
\begin{equation}
w(\lambda, \lambda';\tilde{\mu}_3)=w_0(\lambda;\tilde{\mu}_3)+\frac{1}{2}\left(\frac{ a^2  }{8 \lambda^4}\frac{d w_0}{d\lambda}\right) \lambda'^2,
\end{equation}
which is in the exact form of eqn.\ \eqref{eqn:wdef} with $B(\lambda)=\frac{a^2  }{8 \lambda^4}\frac{d w_0}{d\lambda}$. Thus, if use the rescalings from eqn.\ \eqref{rescale}:
\begin{align}
&Z=\sqrt{-\frac{ B^*}{w_{\lambda \lambda \xi}^*\varepsilon}}\tilde{Z}=\frac{1}{2} \sqrt{-\frac{w^*_{\lambda} a^2}{2 \lambda^4 w^*_{\lambda \lambda \xi}\varepsilon}}\tilde{Z}=0.062...\,  \,\frac{a}{\sqrt{\varepsilon}}\tilde{Z} \notag \\
&\delta \lambda=1.472...\,\sqrt{\varepsilon} \tilde{\delta \lambda} \\ 
&\delta T= \mu \pi a^2 2.911... \,\sqrt{\varepsilon^3}\,\tilde{\delta T}  \notag
\end{align}
we can directly deploy our previous solutions (\ref{firstintegral}-\ref{eqn:eliptic}) to predict  the fibre's behaviour. 

Again following Coleman and Newman, we plot the evolution of a neck under increasing stretch in a fibre of length $L=4 a$ and with $\varepsilon=\tilde{\mu}_3-\tilde{\mu}_3^*=0.5$, which places the homogeneous phases at $\lambda_a=2.547...$ and $\lambda_b=4.484...$. The fibre has a rescaled length $\tilde{L}\approx 45$, which is well into the sub-critical region of instability, and the form of $\lambda(Z)$ is given, analytically, by making the above substitutions into the finite length solution: eqn.\eqref{eqn:finitelengthsolution}. To plot the physical form of the fibre, we then scale the longitudinal direction by $\lambda(Z)$ and the radius by $1/\sqrt{\lambda}$ to conserve volume. The resultant plots are shown in Figure \ref{neck} along the graph of the tension of the system and its bifurcation branch. We note that our solutions can always be shifted longitudinally by half a wavelength, to place either phase in the centre of the rod, leading to energetically-equivalent necking or bulging morphologies. To illustrate this, we show a necked solution during loading and a bulged solution during unloading. The plot clearly shows the subcritical nature of the instability and the hysteretic behaviour of such system. The necking/bulging behaviour qualitatively resembles that originally obtained numerically by Coleman and Newman.

\subsection{Phase Separation in an Elastic Cylinder with Surface Tension}

Soft gel/elastomer fibres  under tension sometimes undergo longitudinal ``beading'' instabilities \cite{matsuo1992patterns,barriere1996peristaltic, mora2010capillarity, ciarletta2012peristaltic,ciarlettanonlinearjmps, ciarlettanonlinearpre, xuan2016finite,Xuan2017}, in which they take on a distinctive ``string-of-sausages'' morphology even though underlying elastic energy is convex. These instabilities are driven by surface tension, forming a solid analogue of the celebrated Plateau---Rayleigh instability in fluids. However, in solids, stability analysis exposes a threshold degree of surface tension required for instability, and indicates that long wavelength modes dominate \cite{barriere1996peristaltic, mora2010capillarity}. Subsequent numerical/weakly-nonlinear studies also revealed that the instability only occurs in a particular stretch-interval, $\lambda_a<\lambda<\lambda_b$, and that, between  $\lambda_a$ and $\lambda_b$, bead amplitude is essentially constant while bead length falls \cite{ciarlettanonlinearpre}. Most recently,  Xuan and Biggins explained this choreography arises because  the instability is a 1D elastic phase separation, with the concavity induced by surface tension \cite{Xuan2017}. Here, we demonstrate that our general near-critical theory applies directly to this instability, yielding a complete analytic treatment in the vicinity of the critical point. 

In this case, the energy of the elastic fibre is the sum of a bulk elastic contribution and a surface energy term. For the elastic part, we take an incompressible neo-Hokean energy density, while the surface energy is simply surface tension, $\gamma$, multiplied by surface area, $A$. However, if the fibre stretches by $\lambda$, incompressibility implies that its radius shrinks by $1/\sqrt{\lambda}$, and therefore its area only rises by $\sqrt{\lambda}$. This leads to a homogeneous energy $\int w_0 \mathrm{d}Z$, with $w_0$ of the form
\begin{equation}
\frac{w_0(\lambda; \Gamma)}{\mu\pi a^2 }=\frac{1}{2}\left(\lambda^2+\frac{2}{\lambda}\right)+2 \Gamma\sqrt{\lambda},
\label{rodsurfaceen}
\end{equation}
where $\Gamma=\gamma/(\mu a)$ is the dimensionless elastocapiliary number. The $\sqrt{\lambda}$ term in the surface energy is a concave function and, beyond a threshold $\Gamma^*$, will introduce a concavity into $w_0$ generating the phase-separation. In this case the critical point can be found analytically, and, as seen in Fig.\ \ref{figurehm}, it lies at $(\lambda^*,\Gamma^*)=(2^{1/3},\sqrt{32})$. 

To regularise the theory, we must consider a gradient term from both the elastic and the surface terms in $w_0$. In general, the surface area $A=\int \frac{d A}{d Z}dZ$ is given by an integral. Expanding the integrand in small $\lambda'$ we get
\begin{align}
\frac{dA}{dZ}=\frac{ 2 \pi a}{\sqrt{\lambda}}\sqrt{\lambda^2+\frac{a^2 \lambda'\,^2}{4\lambda^3}}  = 2 \pi a\left(\sqrt{\lambda}+\frac{1}{2} \frac{a^2 \lambda'^2}{4 \lambda^{9/2}}+...\right),\notag
\end{align}
from which we can directly identify the surface contribution to $B$. We again apply Audoly and Hutchinson's result \cite{Audoly2016}, but only to the elastic part of the homogeneous energy, to get $B_{el}=\frac{a^2 }{8 \lambda^4}\frac{d w_{0el}}{d\lambda}$. Combining these two terms, our overall regularised energy is
\[
w(\lambda, \lambda'; \Gamma)=w_0(\lambda; \Gamma)+\frac{1}{2}\pi a^2 \mu \frac{  a^2  \left(4 \Gamma  \lambda ^{3/2}+\lambda
   ^3-1\right)}{8 \lambda ^6}\lambda'^2 
\]
\begin{figure}[t]
\begin{center}
\includegraphics[width=\columnwidth]{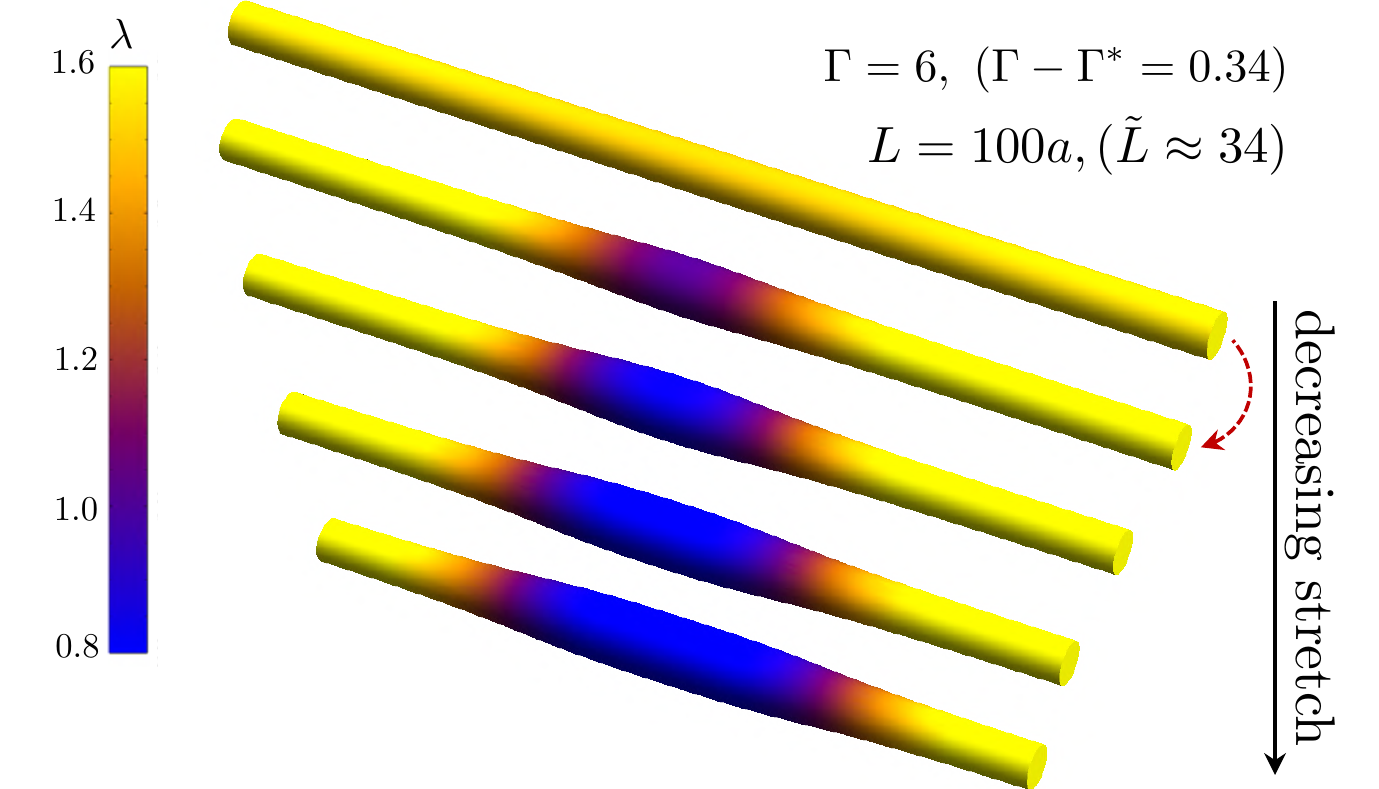}
\end{center}
\caption{ \small Necking evolution for a polymeric fiber near critical point as stretch is decreased (downwards). }
\label{surfacetension}
\end{figure}

This energy is again in the exact form of eqn.\ \eqref{eqn:wdef}, with $\Gamma$ acting as the tuning parameter. We can thus use all our previous solutions (\ref{firstintegral}-\ref{eqn:eliptic}) by making the rescaling substitutions:
\begin{align}
Z&=\frac{\sqrt{33} a}{2^{7/4}\sqrt{\varepsilon}}\tilde{Z},\\
\delta \lambda&=\frac{2^{7/12}}{\sqrt{3}}\sqrt{\varepsilon} \tilde{\delta \lambda},\\
\delta T &=\pi \mu a^2\frac{\tilde{\delta T} \varepsilon ^{3/2}}{2^{11/12} \sqrt{3}},
\end{align}
where $\varepsilon=\Gamma-\Gamma^*$.  For example, inserting these expressions into equation \eqref{tanh}, we see that the expected form for the domain wall in an infinite length system is 
\[
\delta \lambda =\frac{2^{7/12}}{\sqrt{3}}\sqrt{\varepsilon} \tanh\left(\frac{2^{5/4}\sqrt{\varepsilon}}{\sqrt{33} a} Z\right).
\]
This result was previously obtained via small-gradient theory \cite{Xuan2017}.  However, by mapping onto our other results, we are also able to track the whole evolution of phase separation, from the formation of a bead, to its disappearance, in both infinite and finite length systems. In particular, we see that the minimum lengths for instability and subcritical instability respectively are
\[
L_s=\frac{\sqrt{33} a}{2^{7/4}\sqrt{\Gamma-\Gamma^*}}\sqrt{10 \pi}\,,\,\,\,\,\,\,\,\,\ L_h=\frac{\sqrt{33} a}{2^{7/4}\sqrt{\Gamma-\Gamma^*}}2\pi.
\]
Taking, for example, an agar gel cylinder with a radius of $a=240 \mu m$ and subject to $\Gamma=6$  (within the experimental range in \cite{Mora2010}), we can identify that that inhomogenous solutions are entirely suppressed below $L_h= 4.396...\,mm$, and the fibres become become subcritical when the length is greater then $L_s= 6.951...\,mm$.  

To plot the form of the fibre, we again use the fact that the system is locally stretched by a factor $\lambda(Z)$ along the longitudinal direction and, due to impressibility, the thickness is scaled as $1/\sqrt{\lambda}$. We can observe the bulging behaviour during unloading induced by surface tension, as shown in Figure \ref{surfacetension}, for a system of length $L=100a=24 mm$ and with $\Gamma-\Gamma^*\approx0.34$ (corresponding to a $\tilde{L}\approx34$).

\subsection{Phase Separation in Cylindrical Membranes}
\begin{figure}[h]
\begin{center}
\includegraphics[width=\columnwidth]{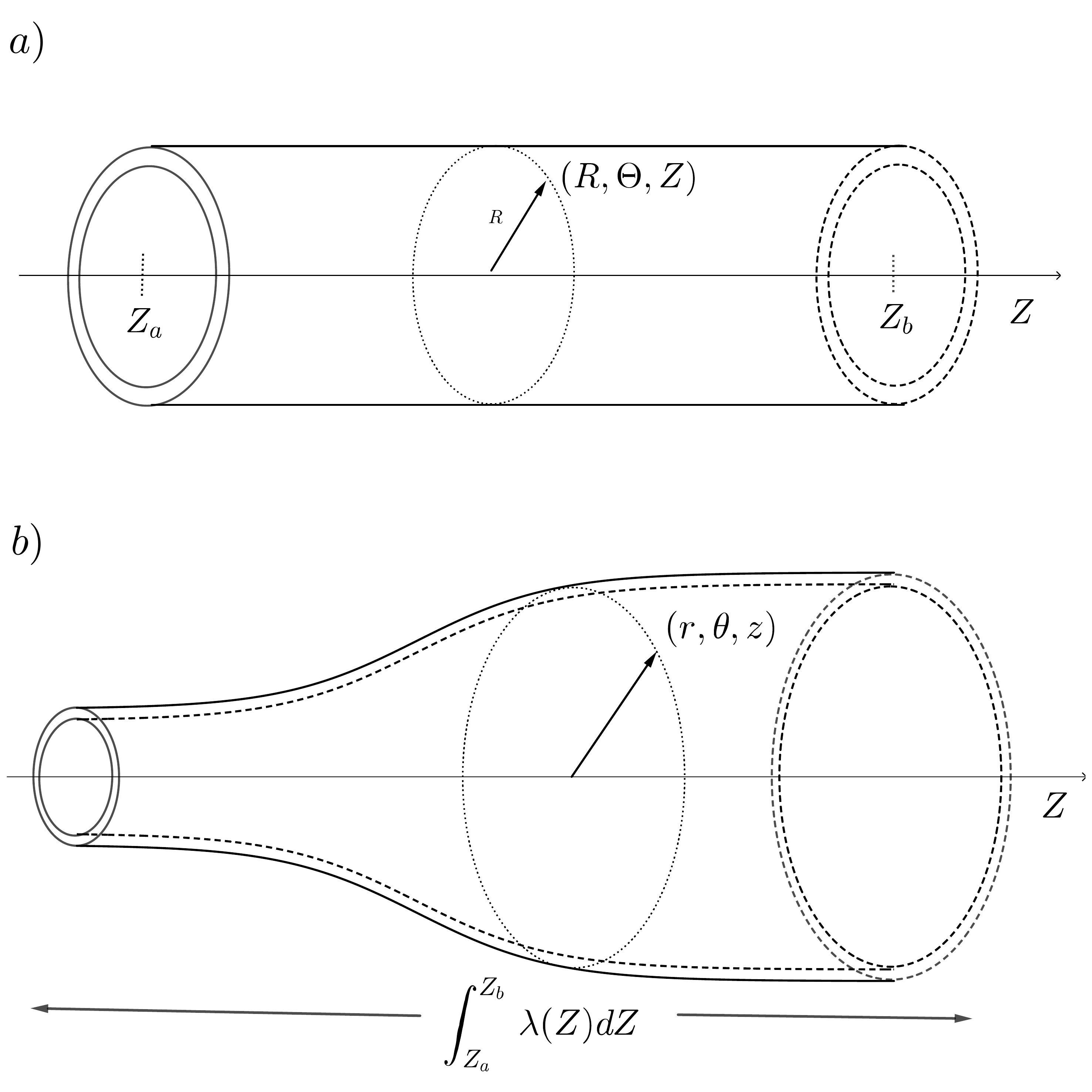}
\caption{\small Schematic of a typical axisymmetric deformation of an initially cylindrical membrane, a), to a phase separated one b).}
\label{bsc}
\end{center}
\end{figure}
Finally, we return to phase separation during the inflation of a cylindrical party balloon. Unlike in the previous examples of rods, where necking/bulging is induced by end to end stretch, phase separation in a party balloon is induced by increase in internal volume. However, the phase separation behaviour can still be traced back to a concavity in the elastic energy as a function of internal volume, $w_0(v)$, which leads to a mechanically unstable region with falling pressure for rising volume. Correspondingly, a homogeneous analysis can be deployed to predict the instability (Conside\'re) points and the two distinct phases, $v_a$ and $v_b$ in an infinite system \cite{Hutchinson1985}. 

To apply our near-critical analysis, we must identify a suitable tuning parameter that can control and remove the concavity. Curiously, if one deploys a simple neo-Hookean elastic model for the balloon, one finds the energy never regains convexity at large volume, so the phase separation has infinite amplitude. In real balloons, $v_b$ is limited by the finite extensibility of polymer chains, which results in dramatic strain stiffening \cite{gent1996new}. Thus, one natural choice for tuning is a constitutive strain stiffening parameter, which is already required to regulate amplitude and can be increased further to eliminate concavity entirely. Alternatively, we will show that concavity can also be removed by applying sufficient tension to the ends of the balloon, allowing any real balloon to be brought into the near-critical regime. 

To resolve phase boundaries one also needs a regularising term penalising gradients in the volume strain. In balloons, such term arises naturally from a full membrane model of the rubber and a recent paper by Lestringant and Audoly \cite{Lestringant2018} has demonstrated how to simplify these expressions for small gradients. A complication arises because the membrane elasticity is naturally expressed in terms of its principle stretches, and a little work is required to eliminate these stretches in favour of the balloon's volume strain. However, this being done, we obtain a  regularized energy for the balloon exactly in the form of eqns.\ \eqref{eqn:wdef}-\eqref{eqg:Gdeff}, with volume playing the role of stretch and pressure playing the role of tension. We may then apply our original solutions to describe the shape evolution of long and short near-critical balloons.

We start by modelling the balloon as as a cylindrical membrane with an undeformed radius $R$, thickness $H\ll R$ and length $L$. If the balloon is then deformed axisymetrically, mapping $(R,\theta,Z)$ to $(r(Z),\theta,z(Z))$ (Figure \ref{bsc}), we can identify the radial stretch as $\eta=\frac{r}{R}$ and the longitudinal stretch as $\lambda=\frac{\partial z}{\partial Z}$. The principle stretches in the membrane are thus $\lambda_1=\eta$ (azimuthal) and $\lambda_2=\sqrt{\lambda^2+R^2 \eta'^2}$ (in membrane longitudinal). If the membrane has a local elastic energy density $W_m(\lambda_1,\lambda_2)$, the total elastic energy of the membrane will then be 
\[
\mathcal{E}=\int 2 \pi R H W_m(\lambda_1, \lambda_2) \mathrm{d}Z\equiv \int \hat{w}(\lambda, \eta,\eta')\mathrm{d}Z,
\label{eqn:membrane}\]
where the energy per unit length of the membrane, $\hat{w}(\lambda, \eta, \eta')$, only depends on its three variables via $\lambda_1$ and $\lambda_2$. This membrane energy already contains $\eta'$, so it is already regularised, and several previous studies have predicted the shape of balloons by numerically minimising energies of this form \cite{Kyriakides1990, Kyriakides1991, Shi1996,meng2014,Lestringant2018}. 

During balloon inflation, the controlled parameter is the volumetric strain $V/V_0=\eta^2 \lambda\equiv v$. (rather than the longitudinal stretch $\lambda$ in previous examples), so we eliminate $\lambda$ in favour of $v$ in the the principle stretches, leading us to a new energy function
\[
\mathcal{E}=\int \hat{w}(v/\eta^2, \eta, \eta')\mathrm{d}Z\equiv \int \tilde{w}(v, \eta,\eta')\mathrm{d}Z.
\]
 We start our treatment of this energy by setting $\eta'=0$, and studying $\tilde{w}_0(v,\eta)=\tilde{w}_0(v,\eta,0)$ which describes homogeneously inflated balloons. To find the balloon shape at  given inflation $v$, we simply need to minimise over $\eta$, setting $\eta=\eta_{min}(v)$ where
\[
\frac{\partial \tilde{w}_0}{\partial \eta}\bigg|_{\eta=\eta_{min}}=0.
\]
This substitution leads us to a homogeneous energy function $w_0(v)=\tilde{w}(v, \eta_{min}(v),0)$ that is only a function of inflation. For example, if the membrane is an incompressible neo-Hookean, its energy function will be $W_m=\frac{1}{2}\mu(\lambda_1^2+\lambda_2^2+1/(\lambda_1 \lambda_2)^2)\equiv \frac{1}{2}\mu I_1$, which leads directly to
\begin{equation}
\eta_{min}=\left(\frac{2 v^4}{v^2+1}\right)^{1/6},\,\,\,\, \,\,\,\,\,\,\,\,  \frac{w_0(v)}{\pi RH \mu}=3\left(\frac{v^2+1}{2v}\right)^{2/3}.\label{eq:etamin}
\end{equation}
We now see clearly the cause of ballooning. In an experiment, we fix $\left<v\right>$, but the above function $w_0(v)$ is concave for $v>\sqrt{\sqrt{21}+4}= 2.929...$ so, if we inflate into this region, the balloon will phase separate into a more inflated and a less inflated region. Correspondingly, the balloon pressure $p= d\mathcal{E}/dV\propto w_0'(v)$ falls in this concave region, indicating mechanical instability. However, with the simple neo-Hooken energy, $w_0(v)$ does not regain convexity at large $v$, so the amplitude of the instability is unbounded. In reality, the instability saturates because, at high strain, the entropic polymer springs in the rubber become fully unwound, dramatically stiffening the rubber. This effect can be included by instead using the Gent constitutive model \cite{GentA.N.1969,Gent2005},
\begin{equation}
W_m(\lambda_1,\lambda_2)=-\frac{1}{2} \mu  J_m\log\left(1- \frac{\lambda_1^2+\lambda_2^2+\frac{1}{\lambda_1^2 \lambda_2^2}-3}{J_m}\right),
\label{Wmballoon}
\end{equation}
which limits to the neo-Hookean model for small strains, but diverges when the strain measure $I_1-3$ approaches the material Gent-parameter $J_m$. Again eliminating $\lambda$ for $v$ then minimising over $\eta$ (requiring $\eta_{min}$ from eqn. \eqref{eq:etamin}), we get the homogeneous energy
\begin{equation}
\frac{w_0(v;J_m)}{\pi R H \mu}=-J_m \log \left(1-\frac{3}{J_m}\left(\left(\frac{v^2+1}{2v}\right)^{2/3}-1\right)\right),
\label{w0balloon}
\end{equation}
which diverges when $\frac{3}{J_m}\left(\left(\frac{v^2+1}{2v}\right)^{2/3}-1\right)=1$, guaranteeing a return to convexity at large $v$ and ensuring a finite amplitude phase separation. 

We next define the homogeneous Gibbs energy (per unit length) of the balloon,
\begin{equation}
G_0(v,p;J_m)=w_0(v;J_m)-p \pi R^2 v.
\label{G0Balloon}
\end{equation}
Here the second term is reminiscent of the term $T \lambda $ in the main section; but now $p$ is introduced as a Lagrange multiplier enforcing a fixed total volume, corresponding to the pressure in the balloon. When $w_0(v;J_m)$ has a concave region, we can define the Maxwell pressure in the same way as we have defined the Maxwell tension in the main section: $p_M$ is the pressure that guarantees $G_0(v,p_M;J_m)$ has two equally deep minima.

\begin{figure}[h]
\begin{center}
\includegraphics[width=\columnwidth]{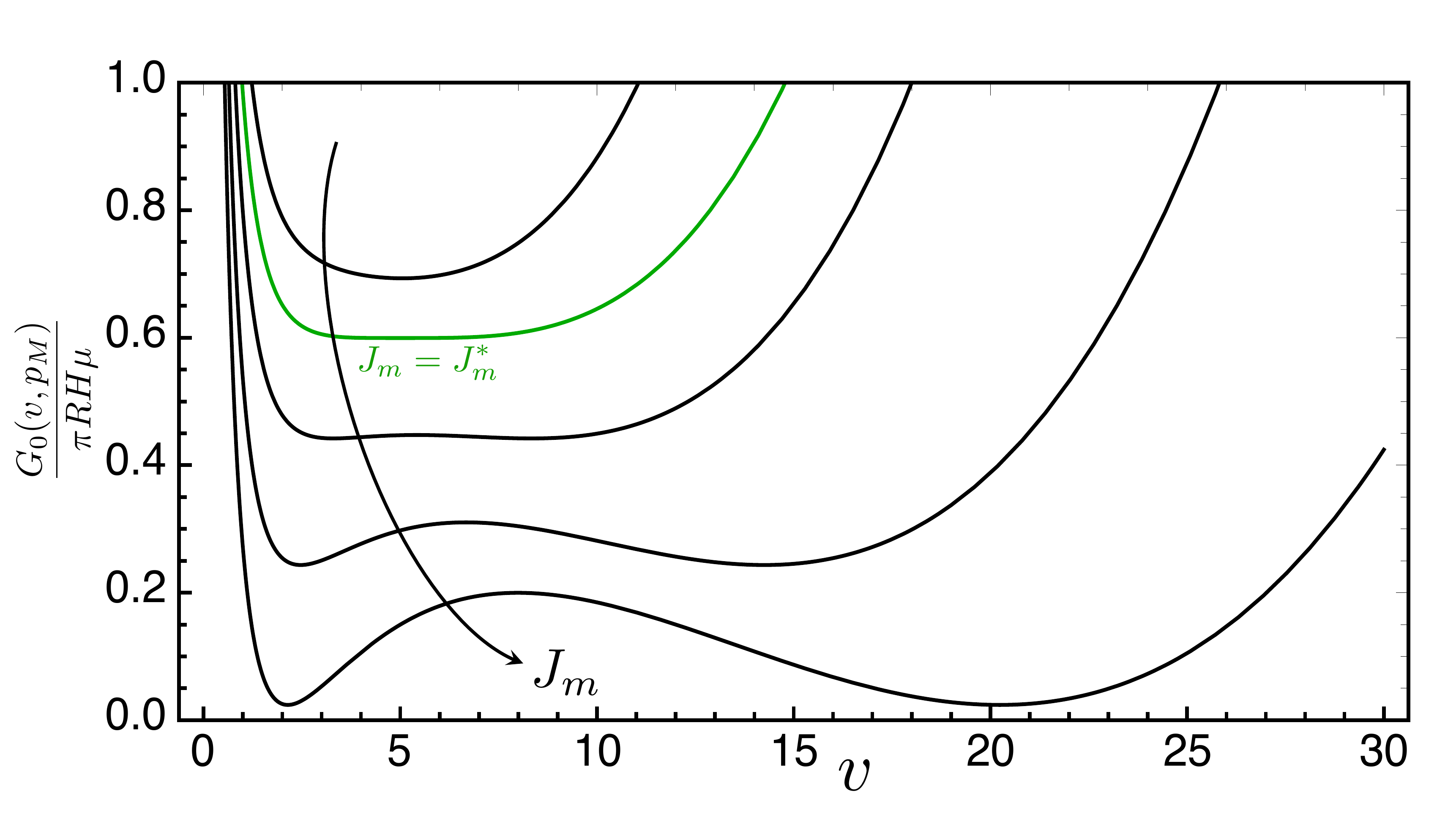}
\caption{\small Plot of the Gibbs homogeneous energy of the balloon at the Maxwell pressure and at different values of the Gent parameter $J_m$. The energy have been offset to better show the developing concave region. As $J_m$ increases, the concave region becomes more prominent and the two minima move further apart.}
\label{be}
\end{center}
\end{figure}

As seen in Figure \ref{be} , the value of the Gent parameter, $J_m$, has a profound effect on the concavity of the Gibbs energy. When $J_m \to \infty$ the Gent system approaches the  neo-Hookean system,  the concavity extends to very high volume strains, and the minima of $G_0(v,p_M;J_m)$ are widely separated. Reducing the  Gent parameter causes $w_0(v)$ to diverge at lower volume strains, bringing the minima of $G_0(v,p_M;J_m)$ closer and closer together, until they merge at $J_m^*=18.23...$, $v^*=4.989...$ at a pressure of $p^*=0.818... \mu H/R$ \cite{Gent2005, meng2014}. Below this value of $J_m$, there is no concavity in $w_0(v)$, and no phase separation. Thus $J_m=\xi$ serves as a parameter controlling the convexity of $w_0(v;J_m)$, and we can conduct a near-critical expansion around $\xi^*=J_m^*$.  

However, before expanding we must move from the homogeneous problem to a regularised in-homogeneous problem, by finding a small-gradient (diffuse interface) form for the elastic energy. We start from $\tilde{w}$, which depends on $(v, \eta,\eta')$ via the two principle stretches, $\lambda_1=\eta$, and $\lambda_2=\sqrt{\left(v/\eta^2\right)^2+(R \eta')^2}$. We follow Lestringant and Audoly \cite{Lestringant2018} and expand the (general) elastic energy in small $(R \eta')^2$. We have that
\[\tilde{w}(v, \eta,\eta')=\tilde{w}_0(v,\eta_0(v))+\frac{\partial\tilde{w}}{\partial\lambda_2}\frac{\partial\lambda_2}{\partial(R \eta')^2}(R \eta')^2+...\,.\]
Evaluation of the derivatives at $(R \eta')^2=0$  guarantees that $\frac{\partial}{\partial\lambda_2}=\eta^2\frac{\partial}{\partial v}$ and $\frac{\partial\lambda_2}{\partial(R \eta')^2}=\frac{\eta^2}{v}$. This  leads to the simple form for the regularised energy for the membrane:
\[
\mathcal{E}=\int \left(\tilde{w}_0(v, \eta)+\frac{1}{2} \frac{\eta^4}{v}\tilde{w}_{0,v} R^2 \eta'^2 +...\right)\mathrm{d}Z,
\label{eqn:membrane}\]
consistent with Lestringant and Audoly. Minimising this energy with respect to $\eta(Z)$ (for a given $v(Z)$ constraining the volume), gives $\eta=\eta_{min}(v)+\mathcal{O}(R^2 \eta'^2)$ (and hence $\eta'=\eta'_{min}(v)v'+\mathcal{O}(R^3 \eta'^3)$), where $\eta_{min}(v)$ is the same function as obtained in the homogeneous case, but now evaluated at the local value of $v$. Inserting these into the above expression and again expanding to $\mathcal{O}(R^2 \eta'^2)$ (remembering that $\eta_{min}$ minimises $\tilde{w}_0(v, \eta)$) we finally get the form we are looking for:
\begin{align}
w(v,v')&= \tilde{w}_0(v, \eta_0(v))+\frac{1}{2} \frac{\eta_{min}(v)^4}{v}\tilde{w}_{0,v} R^2 \eta'_{min}(v)^2 v'^2\notag \\
&\equiv  w_0(v)+\frac{1}{2}B(v) v'^2.
\end{align}
In the particular case of a Gent balloon, $B(v)$ evaluates to:
\[
\frac{B(v)}{\pi R H \mu}=\frac{4 \sqrt[3]{2}  R^2 \left(v+\frac{1}{v}\right)^{2/3} \left(v^6+3 v^4-4\right) J_m}{9 \left(v^2+1\right)^4 \left(2 J_m-3 \sqrt[3]{2}
   \left(v+\frac{1}{v}\right)^{2/3}+6\right)}.
\]
Returning to the Gibbs picture, our  task is now to find the form of $v(Z)$ that minimises:
\[
G(v,v',p;J_m)=w_0(v; J_m)-p \pi R^2 v +\frac{1}{2}B(v;J_m) v'^2.
\]
Making the identifications $v \to \lambda$, $p \pi R^2 \to T$ and $J_m \to \xi$, this is exactly in the form of eqn. \eqref{eqg:Gdeff}, and we can again deploy our entire original analysis. In particular, to use our original solutions (eqns. \eqref{tanh}-\eqref{eqn:eliptic}) in the vicinity of the Gent critical point, $J_m=J_m^*+\varepsilon$, $v=v^*+\delta v$ and $p=p^*+ \delta p$, we simply make the substitutions
\begin{align}
\delta v&= \sqrt{-\frac{6\,w_{vv J_m}^*\varepsilon }{w_{vvvv}^*}} \tilde{\delta \lambda}=1.760...\,\sqrt{J_m-J_m^*}\tilde{\delta \lambda}\\
  Z&= \sqrt{-\frac{ B^*}{w_{vvJ_m}^*\varepsilon}}\tilde{Z}=\frac{4.002...\, R}{\sqrt{J_m-J_m^*}}\tilde{Z}\\
  \delta p &= \frac{1}{\pi R^2}\sqrt{\frac{6 (w_{v v J_m }\varepsilon)^3}{w_{vvvv}^*}} \tilde{\delta T} \\
  &=0.004...\,\frac{ H \mu}{R} (J_m-J_m^*) ^{3/2}\tilde{\delta T} .
  \label{rescaleBalloon}
\end{align}

  \begin{figure}[b]
\begin{center}
\includegraphics[width=\columnwidth]{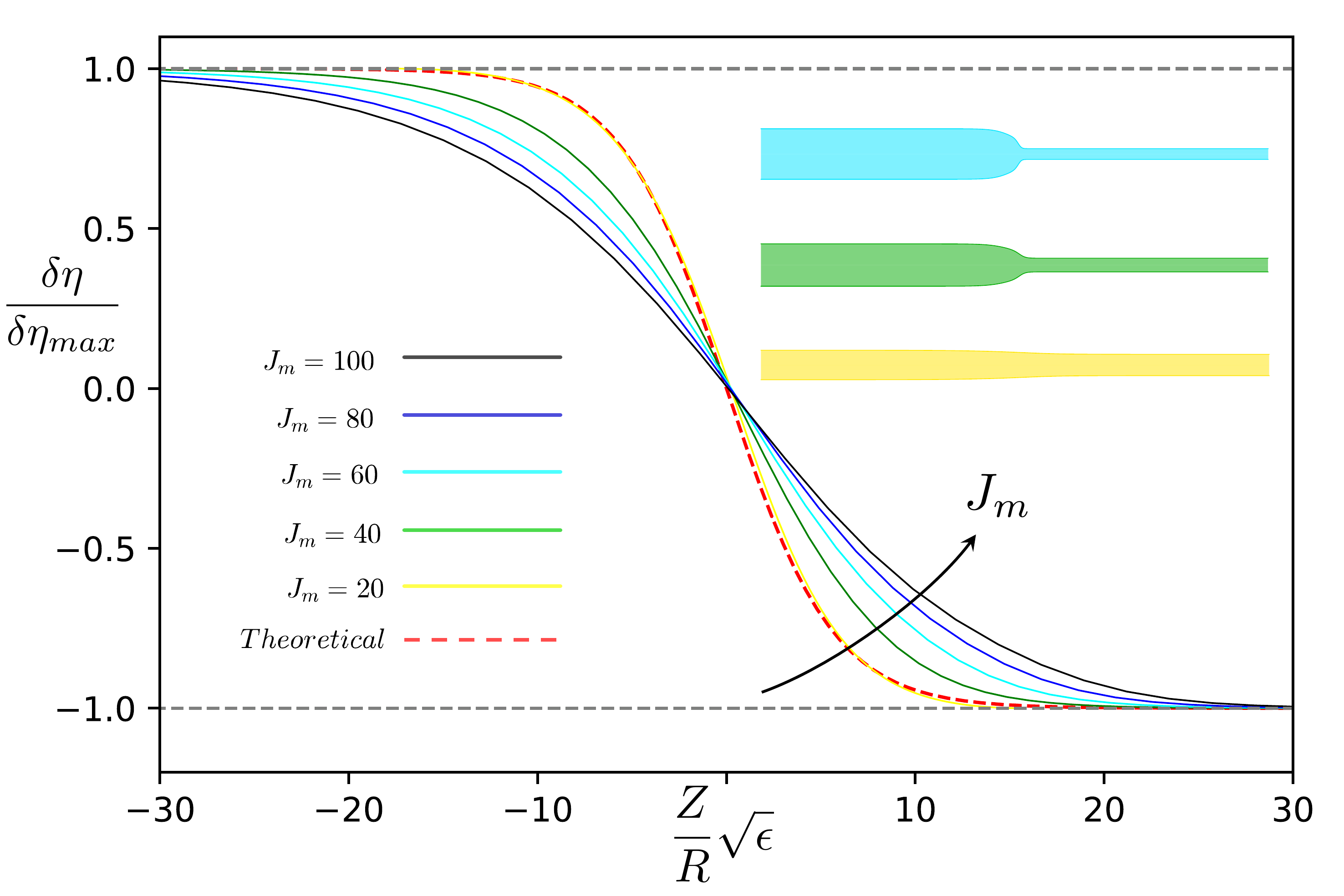}
\caption{\small Plot of numerical results for $\delta \eta(\frac{Z\sqrt{\varepsilon}}{R})$ normalised so that its asymptotes lie at $\pm1$ for different values of $J_m$ from $20$ to $100$. It is clear how the hyper-tangent describes the shape of the interface well also far from the critical point. On the top right, the profiles of balloons with different values of $J_m$ to highlight how the interface is diffuse near the critical point and sharper away from it.  }
\label{comparison}
\end{center}
\end{figure}

These rescalings allow us to deploy our previous solutions to find $\delta v(Z)$ along a bulging balloon. To understand the resultant morphology we also need the corresponding radial and axial stretches. To leading order in small amplitude perturbations, the radial stretch is given by $\eta(Z)=\eta_{min}(v^*)+\eta'_{min}(v^*)\delta v(Z)+O(\delta v^2)$, which evaluates to
\[
\eta(Z)=1.905...+ 0.132... \delta v (Z),
\]
and, substituting this result into $\lambda(Z)=\frac{v}{\eta_{min}(v)^2}$, we find the small amplitude correction to the axial stretch is
\begin{align}
\notag
\lambda(Z)=1.374...+0.084...\delta v(Z).
\end{align}
Having both these function, we can parametrically plot the full final shape of a bulging balloon from our universal solutions.

For example, the $\tanh$ solution for a domain wall in a near-critical Gent balloon has radial stretch profile 
\begin{equation}
\delta \eta(Z)=0.232...\,\sqrt{J_m-J_m^*}\tanh\left(0.176...\, \sqrt{J_m-J_m^*} \frac{Z}{R}\right),
\label{eqn:etatanh}
\end{equation}
which we compare to full finite element simulations of the domain wall at a range of values of $J_m$ (generated by minimising the full Gent membrane energy) in Figure \ref{comparison}. By using  a $Z$ axis scaled by  $\sqrt{J_m-J_m^*}/R$, we demonstrate that all the numerical domain walls agree well with the theoretical $\tanh$, and indeed converge to it as $J_m\to J_m^*$. We emphasise that the un-scaled interface morphology (shown via parametric plots in the top right inset)  is indeed increasingly diffuse near the critical point.

\begin{figure}[t]
\begin{center}
\includegraphics[width=\columnwidth]{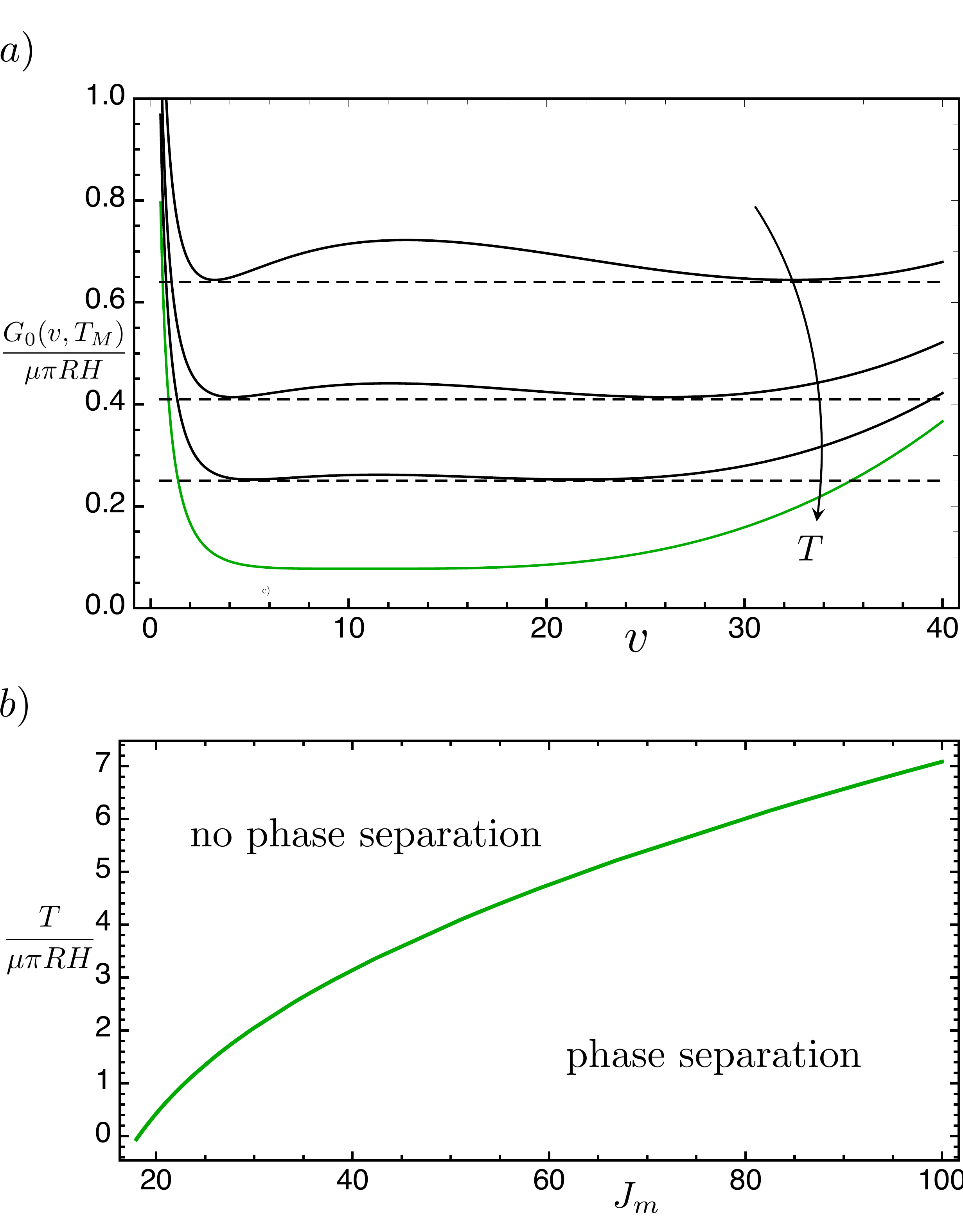}
\caption{\small (a) Plot of the homogeneous energy as a function of $v$ at fixed $J_m=60$ and at different values of $T$. Each energy is shown at the Maxwell pressure and with an offset to highlight the development of the concave region as $T$ is decreased. (b) Plot of the critical tension $T^*$ as a function of the Gent parameter $J_m$. If the system is at any point below the line, its energy is locally concave and it will phase separate. }
\label{JT}
\end{center}
\end{figure}

Our solutions also allow us to plot $p$-$v$ paths (equivalent to the $T$ -$\lambda$ paths in the main section) for inhomogenous solutions, leading to plots which, up to scaling of the axis, would be identical to Figure \ref{f7} a). Previous authors have created analogous plots either experimentally (Fig.\ 14 in \cite{Kyriakides1991}) or by solving the full membrane model computationally (Fig.\ 3a and 4a in \cite{Lestringant2018}). In particular, both these previous papers demonstrated the transition from subcritical to supercritical bulging in short balloons. We are able to derive this length analytically for a near-critical Gent system as  $L_s=\frac{39.76...\,R}{\sqrt{J_m-J_m^*}}$, and also observe that all inhomogeneous solutions are lost below $L_h=\frac{25.15...\, R}{\sqrt{J_m-J_m^*}}$.

In the limit of $\tilde{\delta T} \to\frac{2}{3\sqrt{3}}$ (equivalent to $p$ approaching the Consid\'ere pressure), our infinite length solution (eqn. \eqref{AB}) converges to
\begin{equation}
\tilde{\delta \lambda}=\frac{\sqrt{1+A}}{\sqrt{3}}+\frac{3 A }{2 B}\text{sech} ^2\left(\frac{1}{2}\sqrt{A} (\tilde{Z}-\tilde{Z}_0)\right).
\end{equation}
Pleasingly, this matches the result obtained by several previous authors  for the solution at the onset of bulging in a long balloon without the requirement of near criticality \cite{Fu2008, Pearce2010,Lestringant2018}. In these more complex cases, the solution at the Consid\'ere point nevertheless can be obtained as the amplitude is always low near the point of instability. Regrettably, this solution is of limited interest, since it applies to long balloons in which the instability is sub-critical, meaning the solution is unstable and unobserved. Uniquely, our near-critical treatment provides a full set of analytic solutions, including all those that are stable and observed.

 \begin{figure}[t]
\begin{center}
\includegraphics[width=\columnwidth]{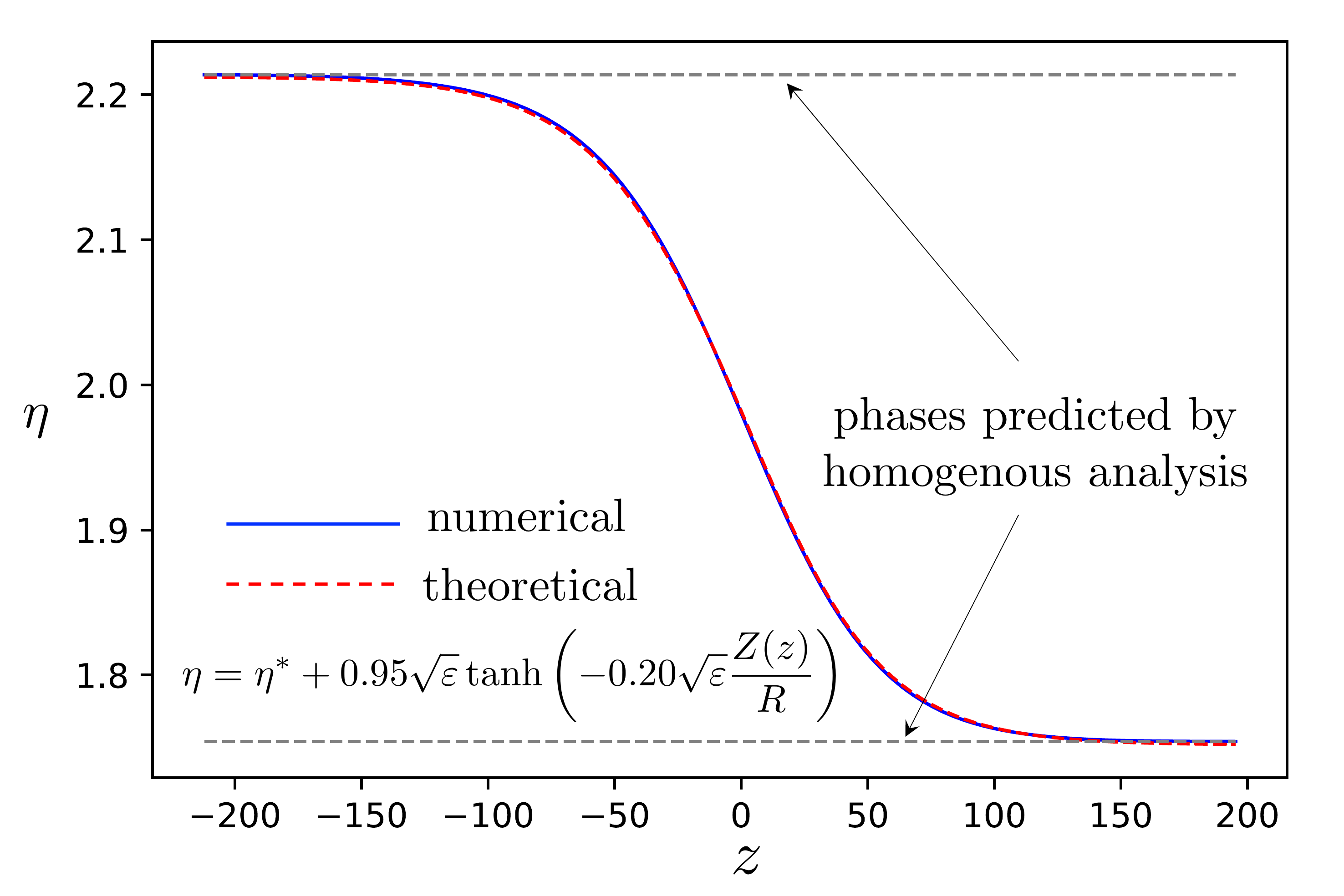}
\caption{\small Comparison between theoretical hyper-tangent (in the final configuration) and numerical results for a system with $J_m=60$ and $\epsilon=\frac{T^*-T}{\mu \pi H R}\approx 0.06$. }
\label{comparisonJT}
\end{center}
\end{figure}

Finally, we consider the introduction of a tension $T$ pulling the two ends of the balloon. Such a tension is absent in a typical party balloon, but is often seen in careful experiments \cite{Kyriakides1990, Kyriakides1991, Wang2019}. As previous authors have noted, \cite{Lestringant2018,Lestringant2020}, including this tension requires the introduction of a term $-T \lambda$ into the energy. Recalling that $\lambda=v/\eta^2$, such a term can easily be included within our framework by the substitution $\tilde{w}(\lambda,\eta, \eta')\to w_{el}(\lambda,\eta, \eta')-T v/\eta^2$, where $w_{el}$ is the elastic contribution analysed previously. In the homogeneous analysis, this new term leads to a more complicated minimisation for $\eta_{min}(v)$ which, for the Gent model, requires the root of a quartic polynomial in $\eta_{min}^2$. Using Mathematica, one may nevertheless solve and thus form a new homogeneous  Gibbs energy $G_0(v,p,J_m;T)$, which now has two tuning parameters $J_m$ and $T$. As illustrated in  Figure \ref{JT} (a), for any $J_m$, one can remove the concavity in $G_0$ by choosing a sufficiently large $T$, allowing one to do a near-critical expansion even in a balloon with $J_m$ far from the critical value treated above. The critical tension required for each choice of $J_m$ is plotted in Figure \ref{JT} (b).

Creating a diffuse-interface (small gradient) regularised energy with tension proceeds exactly as before, except $B(v;T)$ should be computed only from $w_{el}$, as there are no gradient terms in the new $T v/\eta^2$ term. In Figure \ref{comparisonJT}, we show a comparison between the shape of a balloon under tension obtained numerically from the full non-linear membrane theory and the theoretical result derived using perturbations of the tension parameter $T$, for a system with $J_m=60$. The critical tension is $T^*=4.758...\,\mu \pi H R$ and we choose $T=4.7\mu \pi H R$  with $\varepsilon=\frac{T^*-T}{\mu \pi H R}\approx0.06$; yielding excellent agreement.

Again, up to the re-scaling of the axes using eqn. \eqref{rescale} at a fixed $J_m^*$ with $\lambda\to v$, $T \to \pi R^2 p$ and $\xi \to \frac{T}{\mu \pi H R}$, the shape evolution of the profile of a bulging balloon under tension is shown in Figure \ref{figsolitonsol} (c)  for a long system and in Figure \ref{fl}  for a finite length one. We would like to highlight the striking similarity between Figure \ref{figsolitonsol} (c) and Figure 6 in  \cite{Pearce2010} as well as Figure 14 and 21 in \cite{Wang2019}, picturing shapes for a bulging balloons under tension far from the critical point obtained numerically in the first case, and experimentally in the latter two images. 

For the purposes of this discussion, we have chosen to perform our balloon analysis using $v$ and $\eta$ as variables, rather than $\lambda$ and $\eta$. Our  approach has had the advantage that $v$ is the naturally constrained variable during inflation, so using $v$ exposes the concavity of the homogeneous elastic energy. However, the price paid is that the balloon energy is simpler when described using $\eta$ and $\lambda$, as becomes particularly clear when tension is included. Alternatively, one could work directly with $\lambda$ and $\eta$,  expand in both  around the critical point, and then minimise over both fields after expansion. This approach ultimately leads to the same set of near-critical solutions, and generalises more easily to problems with more fields and constraints. 

\section{Discussion and Conclusion}
We have conducted an extended analysis of a 1D elastic system subject to  longitudinal phase separation, such as necking in elastic fibres or bulging in cylindrical party balloons. Our main conclusion is that, in the vicinity of a critical point, where the amplitude of phase separation vanishes, such systems are governed by a universal energy. We can minimise this energy analytically to find the exact solutions to the non linear problem and hence the resultant shapes of finite and infinite length systems, and predict their full shape evolution and hysteresis loop. 

Importantly, at a distance $\varepsilon$ from the critical point, the instability amplitude is very small, $\mathcal{O}(\sqrt{\varepsilon})$, while the characteristic length-scale for variation along the system (e.g. a domain wall) diverges as $\mathcal{O}(1/\sqrt{\varepsilon})$. Thus, in the  critical region, it is self consistent to assume  both slow variations (diffuse interfaces) and low amplitude. This contrasts with previous work which has assumed small gradients while allowing large variations in amplitude. Our treatment also identifies and locates a transition from sub-critical to super-critical instability below an analytic threshold length, and the vanishing of all inhomogeneous solutions below a second analytic length. 

The near-critical elastic model has strong antecedents. The tanh interfacial solution for an infinite-length system with a symmetric potential is familiar from many contexts, including near-critical fluid-vapour interfaces \cite{Langer,Cahn1958} and magnetic domain walls. However in the context of elastic instabilities (unlike in typical thermodynamic contexts) it is natural to investigate inhomogeneous solutions in finite length systems with asymmetric potentials, leading to the large and unfamiliar set of solutions discussed here. Indeed, similar families of solutions has been previously discussed within the elasticity field  \cite{triantafyllidis1986gradient,Triantafyllidis1993}, but previous authors have focused on general (non critical) elastic energies. In these cases, provided the regularisation is a simple quadratic, one can (following van-der-Waals \cite{VanderWaals1979})  generate  the corresponding differential equation and conduct an analytic first integral, but a numerical calculation is required to conduct a second integral and plot the (system-specific)  solutions. Our work thus exploits the simplicity and universality of a near-critical system to elucidate the rich behaviour of an elastic system.

Our elastic model may be simple and low amplitude, but it captures far more than a typical linear stability analysis: in particular, it tracks the full non-linear evolution of the inhomogeneous solutions, including their  diffuse onset (as captured in linear stability analysis), their rapid localisation into an interface, the interface's propagation along the rod, and its ultimate delocalisation as the rod returns to homogeneity. Our model thus provides a playground for studying localisation in elastic systems: for example, it could underpin dynamical studies on the growth of necks and elastic coarsening.  Similarly, extending the model to include imperfections \cite{Triantafyllidis1993} may offer insights into how they  stabilise finite wavelength modes, as observed in the capillarity driven beading of gel cylinders \cite{mora2010capillarity}. Looking beyond 1D, a similar ``near-critical'' approach might also capture the highly topical phenomena of buckling localisation \cite{audoly2020localization} and imperfection sensitivity  in compressed shells.

\acknowledgements{A.G.\ thanks the EPSRC for funding, project 2108804. JSB.\ is supported by a UKRI Future Leaders Fellowship MR/S017186/1}

\bibliography{pslibrary.bib}

\section{Appendix A: Expansion of the Bifurcated Branch}
\textbf{Linear stability analysis}\\

To find the bifurcation point, let us consider $(\tilde{\delta \lambda}_A,\tilde{\delta T}_A)$  corresponding to point $A$ in Figure \ref{f7} b).
First, via \eqref{diffeq}, we have that $\tilde{\delta T}_A=-\tilde{\delta \lambda}_A+\tilde{\delta \lambda}_A^3$. 
We next consider a small sinusoidal perturbation to this homogeneous state
$$\tilde{\delta \lambda}=\tilde{\delta \lambda}_A+\epsilon \cos(k \tilde{Z}),\,\,\,\,\,\, \tilde{\delta T}=\tilde{\delta T}_A+\epsilon \tilde{\delta T}_1.$$ 
Substituting these into \eqref{diffeq} and and linearizing in $\epsilon$, we see that $\tilde{\delta T}_1=0$ and the mode shape requires  $k=\sqrt{1-3(\tilde{\delta \lambda}_A)^2}$.  
The linearized boundary conditions \eqref{b.c.} require $k=2\pi/ \tilde{L}$ for the first unstable mode, which we can solve for the bifurcation threshold  as a function of $\tilde{L}$,
\begin{equation}
\tilde{\delta \lambda}_A=-\frac{\sqrt{\tilde{L}^2-4 \pi ^2}}{\sqrt{3}\tilde{L}}.
\label{dlA}
\end{equation}
Since, by symmetry, the instability corresponding to point $D$ in Figure \ref{f7} a) satisfies $(\tilde{\delta \lambda}_A,\tilde{\delta T}_A)=(-\tilde{\delta \lambda}_D,-\tilde{\delta T}_D)$, when $\tilde{\delta \lambda}_A=\tilde{\delta \lambda}_D=0$ the two instability points merge and no inhomogeneous solutions are observed; this happens at the length $\tilde{L}_h=2 \pi$.\\

\textbf{Koiter analysis}\\

To find the length $\tilde{L}_s$ that identifies the transition from sub- to super-critical behaviour in the bifurcation, we need to find the gradient at the onset of instability. We use a Koiter-like analysis \cite{koiter1970stability} and we expand our solution in the vicinity of the bifurcation point, using the amplitude ($\epsilon$) of the $\cos(k Z)$ Fourier mode as a small parameter. To satisfy the boundary conditions, we only expect cosine functions to appear in $\tilde{\delta\lambda}(\tilde{Z})$, leading us to the proposed solution:
\begin{align}
\tilde{\delta \lambda}(\tilde{Z})=\tilde{\delta\lambda}_A&+\epsilon \cos(k \tilde{Z})+\epsilon^2(c_1+c_2 \cos(2k \tilde{Z}))\\
&+\epsilon^3(c_3+c_4\cos(2k \tilde{Z})+c_5\cos(3k \tilde{Z}))+...\notag\\
\tilde{\delta T}=\tilde{\delta T}_A&+\epsilon^2 \tilde{\delta T}_2+\epsilon^3 \tilde{\delta T}_3+...\,.
\end{align}
The zeroth and first order were solved in the linear stability analysis. The second order bulk equation is now solved provided 
\begin{equation}
c_1=\frac{2 \tilde{\delta T}_2-3 \tilde{\delta \lambda}_A}{2 \left(3 \tilde{\delta \lambda}_A^2-1\right)}\,\,\,\,\,,\,\,\,\,\,\,c_2=\frac{\tilde{\delta \lambda}_A}{6\tilde{\delta \lambda}_A^2-2},
\end{equation}
while, to fix $\delta T_2$, we need to go to third order to obtain
\begin{equation}
\delta T_2=\frac{1+7 \tilde{\delta \lambda}_A^2}{8 \tilde{\delta \lambda}_A}.
\end{equation}

To find the initial gradient we first note that the average stretch of the solution near the bifurcation point can be obtained using the integral in eqn. \eqref{totstretch}. Since the cosine functions do not contribute to this integral, we obtain that
\begin{align}
<\lambda>&=\tilde{\delta\lambda}_A+ \epsilon^2 c_1+...
\end{align}
Finally, using that the initial gradient of the bifurcation is $m_A=\frac{d(\tilde{\delta T})}{d(<\lambda>)}$ we obtain that,
\begin{equation}
m_A=\frac{\tilde{\delta T}_2}{c_1}=\frac{20 \pi ^2 \tilde{L}^2-56 \pi ^4}{\tilde{L}^4-10 \pi ^2 \tilde{L}^2},
\end{equation}
where the second equality also requires substitution from eqn.\ \eqref{dlA}. The length $\tilde{L}_s$ is obtained by solving $1/m_A=0$, which is equivalent to the point where the gradient changes sign, at $\tilde{L}_s=\sqrt{10} \pi$.

\section{Appendix B: Details on Numerical Simulations for Cylindrical Membranes}
\begin{figure}[b]
\begin{center}
\includegraphics[width=7 cm]{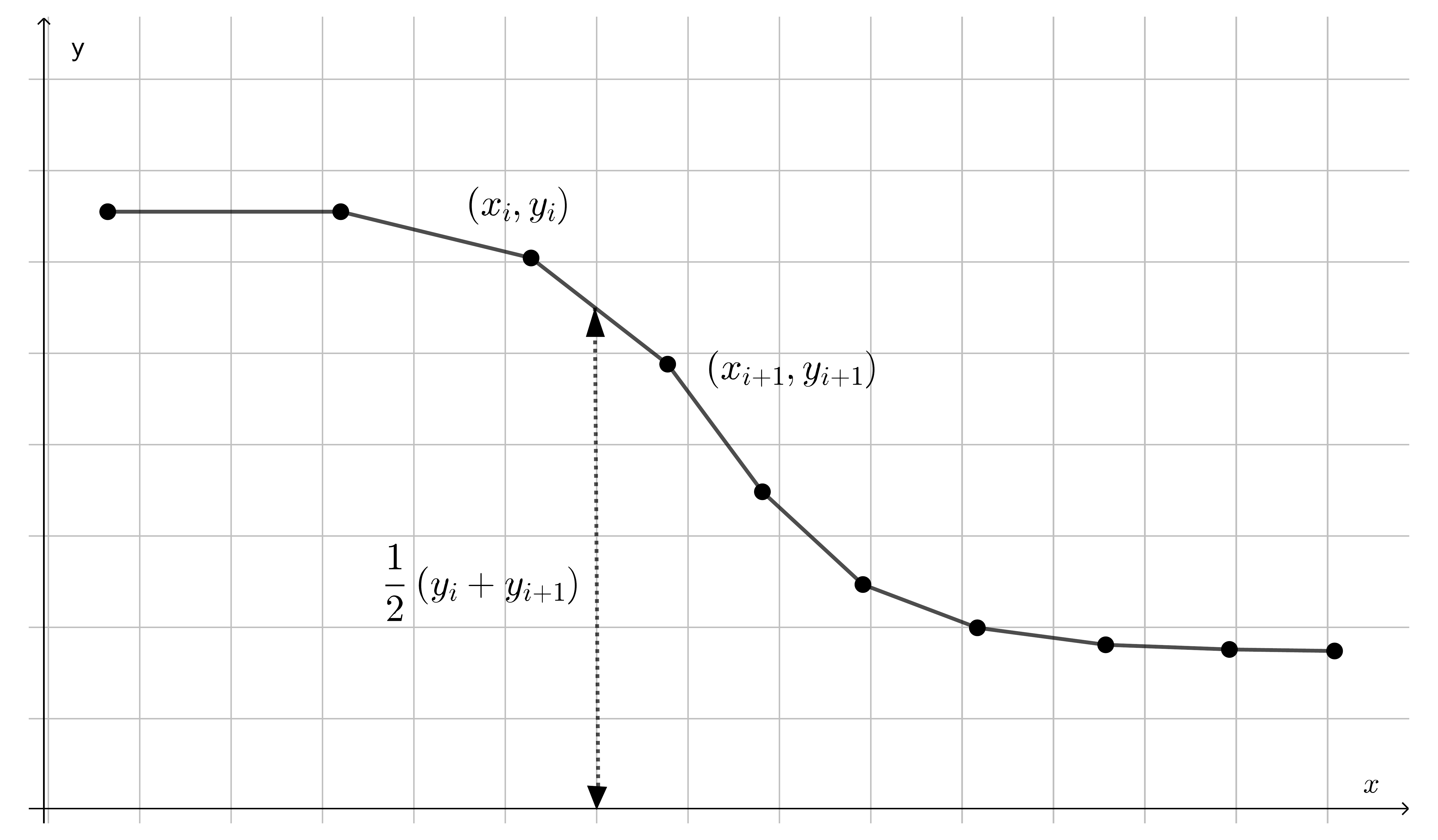}
\end{center}
\caption{\small\label{numerical} Schematics of sections and masses used for numerical calculations of the profile of the domain wall.}
\end{figure}
The numerical results have been obtained in two ways. To find the shape of the domain wall we discretised the system by cutting the $Z$ line in $n$  equally spaced sections of length $D=L/n$ at the end of which we have a mass $m_i$ with position $(x_i,y_i)$. We then assume the energy of the $i^{th}$ section is given by
\[\mathcal{E}_i=D\left(W_m(\lambda_{1i},\lambda_{2i})-p \lambda_1i^2 \lambda_i\right),\]
where $\lambda_{1i}$ is equal to the radial stretch $\eta_i$ and the membrane energy is the one appearing in equation \eqref{Wmballoon}. As shown in Figure \ref{numerical}, the radial stretch is given by its value at the midpoint of the section, $\lambda_{1i}=\eta_i=\frac{1}{2}\left(y_i+y_{{i-1}}\right)$ and the second principal (in membrane) stretch,  $\lambda_{2i}$, is given by $\lambda_{2i}=\frac{(y_i-y_{i-1})^2+(x_i-x_{i-1})^2}{D}$. Note that the longitudinal stretch is simply $\lambda_i=\frac{x_i-x_{i-1}}{D}$. 
The total energy, which is the sum of the energies of each section, is a function of the position of the $n$ masses. We minimize the energy of this function over the $2n$ variables using the \textit{scipy.optimize} library on \textit{python}, constraining the volume and feeding a suitable guess to improve efficiency and avoid any trivial solution. This is analogous to the technique used in \cite{meng2014}. For our simulations, we used 100 sections (101 masses) as any greater number proved to be unnecessary.  To check for stability, we wrote a one dimensional dynamic code where each mass is subject to the (elastic) force of the the two adjacent sections (obtained easily from $\mathcal{E}_i$) as well as a vertical force arising from the internal pressure.

\end{document}